\documentclass[twocolumn, prb, superscriptaddress]{revtex4}
\usepackage{color, graphicx}
\usepackage{amssymb}
\usepackage{gensymb}

\begin{document}

\title{Strong anisotropy in nearly ideal-tetrahedral superconducting FeS single crystals}

\author{Christopher K. H. Borg}
\affiliation{Department of Chemistry and Biochemistry, University of Maryland, College Park, MD 20742}
\author{Xiuquan Zhou}
\affiliation{Department of Chemistry and Biochemistry, University of Maryland, College Park, MD 20742}
\author{Christopher Eckberg}
\affiliation{Department of Physics, University of Maryland, College Park, MD 20742}
\affiliation{Center for Nanophysics and Advanced Materials, University of Maryland, College Park, MD 20742}
\author{Daniel J. Campbell}
\affiliation{Department of Physics, University of Maryland, College Park, MD 20742}
\affiliation{Center for Nanophysics and Advanced Materials, University of Maryland, College Park, MD 20742}
\author{Shanta R. Saha}
\affiliation{Department of Physics, University of Maryland, College Park, MD 20742}
\affiliation{Center for Nanophysics and Advanced Materials, University of Maryland, College Park, MD 20742}
\author{Johnpierre Paglione}
\affiliation{Department of Physics, University of Maryland, College Park, MD 20742}
\affiliation{Center for Nanophysics and Advanced Materials, University of Maryland, College Park, MD 20742}
\author{Efrain E. Rodriguez*}
\affiliation{Department of Chemistry and Biochemistry, University of Maryland, College Park, MD 20742}
\affiliation{Center for Nanophysics and Advanced Materials, University of Maryland, College Park, MD 20742}

\begin{abstract}
We report the novel preparation of single crystals of tetragonal iron sulfide, FeS, which exhibits a nearly ideal tetrahedral geometry with S--Fe--S bond angles of 110.2(2) $^\circ$ and 108.1(2) $^\circ$. Grown via hydrothermal de-intercalation of K${_x}$Fe${_{2-y}}$S${_2}$ crystals under basic and reducing conditions, the silver, plate-like crystals of FeS remain stable up to 200 $^\circ$C under air and 250 $^\circ$C under inert conditions, even though the mineral ``mackinawite'' (FeS) is known to be metastable.  FeS single crystals exhibit a superconducting state below $T_c=4$ K as determined by electrical resistivity, magnetic susceptibility, and heat capacity measurements, confirming the presence of a bulk superconducting state. Normal state measurements yield an electronic specific heat of 5~mJ/mol-K$^2$, and paramagnetic, metallic behavior with a low residual resistivity of 250~$\mu\Omega\cdot$cm.  Magnetoresistance measurements performed as a function of magnetic field angle tilted toward both transverse and longitudinal orientations with respect to the applied current reveal remarkable two-dimensional behavior. This is paralleled in the superconducting state, which exhibits the largest known upper critical field $H_{c2}$ anisotropy of all iron-based superconductors, with  $H_{c2}^{||ab}(0) / H_{c2}^{||c}(0)=$(2.75~T)/(0.275~T)=10. Comparisons to theoretical models for 2D and anisotropic-3D superconductors, however, suggest that FeS is the latter case with a large effective mass anisotropy.  We place FeS in context to other closely related iron-based superconductors and discuss the role of structural parameters such as anion height on superconductivity.

\end{abstract}

\maketitle


\section{Introduction}

While the field of iron-based superconductors has focused primarily on selenides, tellurides, and arsenides,\cite{Paglione2010, Johnston2010, Ivanovskii2011} recent developments show that sulfides are a possible new avenue for high-$T_c$ superconductors.  The first iron-sulfide superconductor, BaFe$_{2}$S$_{3}$, has been reported to have a superconducting critical temperature ($T_{c}$) = 14 K at 11 GPa.\cite{Takahashi2015}  An even simpler sulfide, H$_2$S, under high pressure (90 GPa), has been found to exhibit superconductivity as high as 203 K, which is the highest reported $T_c$ thus far.\cite{Drozdov2015} Sulfides in general therefore merit closer inspection for exploring high temperature superconductivity, and iron sulfides in particular could point the way towards new superconducting compounds.

\begin{figure}[t!]
\centering
  \includegraphics[width=0.95\columnwidth]{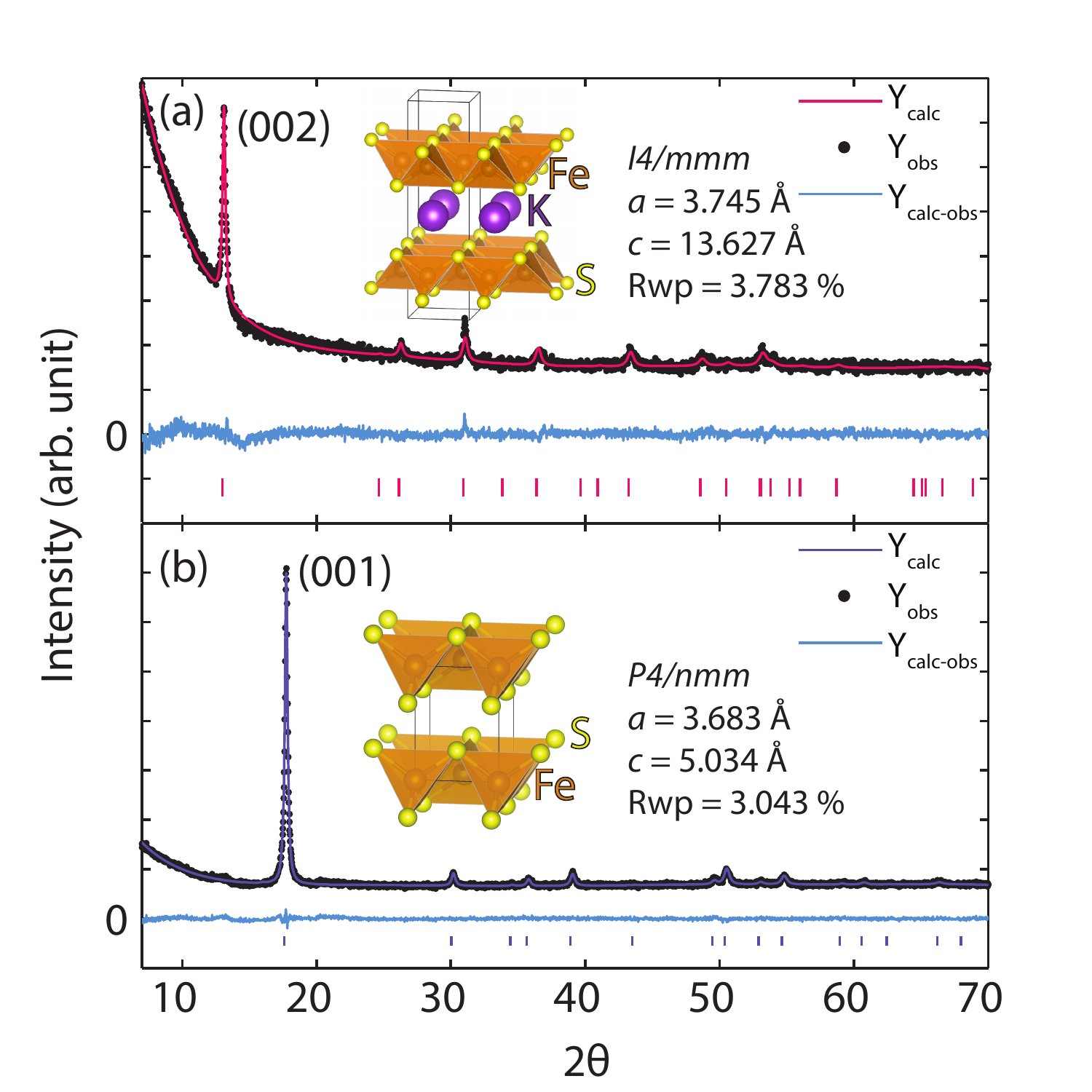}
  \caption{Rietveld refinement with XRD powder diffraction on ground single crystal samples. (a) Refinement of K${_x}$Fe${_{2-y}}$S${_2}$ template's body-centered tetragonal structure ($I4/mmm$). (b) Refinement of the FeS product's structure with a primitive tetragonal model ($P4/nmm$). Fe (orange) ions are tetrahedrally coordinated to S (yellow) anions, and the K (purple) cations are located between two FeS layers. Tick marks corresponding to their respective phase are shown below the difference curve.}
  \label{fgr_XRD}
\end{figure}

Recently, Lai \textit{{\it et al.}} found that the simple binary compound, FeS, in its tetragonal polymorph known as mackinawite is a superconductor with a $T_c =5$ K.\cite{Lai2015}  Similar to the superconducting $\beta$-form of iron selenide, mackinawite also adopts the anti-PbO structure where FeS$_4$ tetrahedra edge-share to form two-dimensional (2D) layers (Figure~\ref{fgr_XRD}b inset).\cite{Kouvo1963, Bertaut1965, EvansJr1964}  Unlike its heavier analogues, FeSe and FeTe, however, mackinawite is metastable and therefore cannot be synthesized from their respective elements using solid state methods, unless it is alloyed with significant amounts of Co, Ni or Cu.\cite{Lennie1995, Rickard2007}  Due to the thermodynamic limitations in its preparation, single crystal growth of mackinawite is a challenge.  Growing single crystals of FeS is imperative, however, towards understanding its true physical properties.

Before the report by Lai \textit{{\it et al.}} on superconductivity, several studies had found FeS to be a ferrimagnetic semiconductor. \cite{Denholme2014, Sines2012}  The conflicting reports on the properties of polycrystalline FeS by different groups may be due to impurities not observed through powder X-ray diffraction, especially since iron provides a high background from fluorescence with Cu K-$\alpha$ radiation.  Powder FeS samples prepared through aqueous methods may form small crystallites as indicated by the broad Bragg reflections in the diffraction patterns of past studies.\cite{Dutta2012}  The small particle size and polycrystalline nature of these samples impede accurate electrical resistivity and magnetization measurements due to grain boundary effects and the facile oxidation of surfaces of small particles.\cite{Denholme2014, Bertaut1965} Despite their ground-breaking work on polycrystalline FeS, Lai \textit{{\it et al.}} also called for high quality single crystal data for definitive determination of the physical properties of FeS.

We report a method for the preparation of high quality single crystals of mackinawite FeS.  Since FeS is metastable,\cite{Lennie1995a, Malasics2012} single crystal growth through slow cooling of a melt is not possible.  In the case of FeSe$_{1-y}$Te$_y$\cite{Bao2009, Liu2010, Bhatia2011} and Fe$_{1+x}$Te,\cite{Rodriguez2011_a, Stock2011} large single crystals were grown through Bridgeman techniques allowing detailed transport and spectroscopic experiments.  For FeSe, which has limited window of phase stability, chemical vapor transport methods at elevated temperatures is the only technique that has been reported.\cite{Hara2010, Boehmer2015}  We present a general technique for the de-intercalation of the ternary phase  K${_x}$Fe${_{2-y}}$S${_2}$ Figure~\ref{fgr_XRD}a inset), which melts congruently and can therefore be prepared in single crystal form.\cite{Hu2011, Luo2011} We link how studying the materials chemistry of layered iron sulfides is key to discovering the underlying physics in new superconductors such as mackinawite FeS.

\section{Experimental}

\subsection{Hydrothermal synthesis of FeS single crystals}
In this work, superconducting FeS single crystals were prepared by de-intercalation of potassium cations from K${_x}$Fe${_{2-y}}$S${_2}$ ($x \approx 0.8$, $y \approx 0.4$) single crystals under hydrothermal condition. The growth of K${_x}$Fe${_{2-y}}$S${_2}$ single crystals was modified by the method described by Lei \textit{{\it et al.}} \cite{Lei2011}  For a typical reaction, 1.00 g (11.4 mmol) of hexagonal FeS powder (Alfa Aesar, 99.9\%) was mixed with 0.18 g (4.5 mmol) of potassium metal (Alfa Aesar, 99\%) to match the nominal composition of K${_{0.8}}$Fe${_{2}}$S${_2}$. The mixture was loaded in a quartz ampoule inside an argon-filled glovebox, and the ampoule was flame sealed under vacuum (10$^{-3}$ Torr). In order to avoid oxidation of the sample due to the potassium-induced corrosion of quartz, the sample containing ampoule was sealed in a larger ampoule under vacuum (10$^{-3}$ Torr). 

For crystal growth of  K${_x}$Fe${_{2-y}}$S${_2}$, the mixture was heated to 1000 $^{\circ}$C over 10 hours and held at 1000 $^{\circ}$C for 3 hours to form a homogeneous melt. Subsequently, the melt was slowly cooled at a rate of 6 $^{\circ}$C/hour to 650 $^{\circ}$C to allow crystal growth. After cooling to room temperature, K${_x}$Fe${_{2-y}}$S${_2}$ single crystals approximately 3 mm -- 8 mm in diameter and approximately 0.1 mm in thickness were recovered. 

For the preparation of FeS single crystals, the K${_x}$Fe${_{2-y}}$S${_2}$ precursor (0.2 g - 0.4 g), 0.28 g (5 mmol) Fe powder (Alfa Aesar, 99.9\%), 0.84 g (5 mmol)  Na$_{2}$S ${\cdot}$ ${5}$H$_{2}$O (dried from Na$_{2}$S ${\cdot}$ ${9}$H$_{2}$O, Sigma-Aldrich, 98\%) and 0.20 g (5 mmol) NaOH (Sigma-Aldrich, 98\%) were added to 10 mL water. The mixture was placed in a Teflon-lined stainless steel autoclave at 120 $^{\circ}$C for 3-4 days. Silver colored FeS single crystals were recovered by washing away excess powder with water and drying under vacuum overnight. Samples prepared in the absence of excess iron powder were not superconducting, which could be due to either oxidation of the iron or vacancy formation in the FeS layer.   In the crystallographic studies of layered iron selenide analogues such as FeSe\cite{McQueen2009} and (Li$_x$Fe$_{1-x}$OH)FeSe,\cite{Sun2015} iron vacancy formation is implicated in the loss of superconducting properties.

\subsection{X-ray diffraction and thermal stability analysis}
Initial powder X-ray diffraction (XRD) data were collected using a Bruker D8 X-ray diffractometer with Cu K${\alpha}$ radiation, ${\lambda}$ = 1.5418 \AA \,(step size = 0.025$^{\circ}$, with 2${\theta}$ ranging from 7$^{\circ}$ - 90$^{\circ}$). Temperature dependent X-ray diffraction on ground single crystals was performed using a Bruker C2 diffractometer with a Vantec500 2D detector, ${\lambda}$ = 1.5418 \AA \, (step size = 0.05$^{\circ}$, with 2${\theta}$ ranging from 11$^{\circ}$ - 80$^{\circ}$). The sample was heated using an Anton Paar DHS 1100 graphite-dome hot stage. Rietveld refinements were carried out using TOPAS software.

Differential scanning calorimetry (DSC) was conducted on a Mettler-Toledo TGA/DSC 3+ thermogravimetric analyzer with high temperature furnace. Samples were heated from room temperature to 800 $^{\circ}$C. 


\subsection{Magnetic susceptibility, electrical transport and heat capacity}

Magnetic susceptibility measurements were performed using a Quantum Design Magnetic Properties Measurement System (MPMS). Both field-cooled (FC) and zero field-cooled (ZFC) measurements were taken from 2 K to 300 K in direct current mode with an applied magnetic field of 10 Oe -- 30 Oe. Hysteresis measurements were carried out at 2 K with $H = {\pm}$7~T. Magnetic susceptibility measurements under hydrostatic pressure were performed using a BeCu piston-cylinder clamp cell employing n-pentane:isoamyl alcohol as a pressure-transmitting medium. Pressures produced on the single crystal sample at low temperatures were calibrated by measuring the Meissner effect of a small piece of Pb, placed in the pressure cell. The known pressure dependences of the superconducting transition temperature of Pb \cite{Smith1967} were used for this purpose.

Electrical transport measurements were performed on a 14 T Quantum Design Dynacool Physical Properties Measurement System (PPMS). Single crystal samples were mounted on a rotator AC transport sample board and measured using the electrical transport option, applying currents between 0.1-0.5~mA and frequencies near 10~Hz. 

Heat capacity measurements were performed in a 14 T Quantum Design Dynacool PPMS System. The single-crystal sample of mass 2.9~mg was measured using the relaxation method with field applied perpendicular to the basal plane.

\section{Results: Synthesis, thermal stability and structural characterization}

\subsection{Single crystal preparation by reductive de-intercalation}

Our strategy for preparing single crystals of a metastable phase can be summarized as crystal-to-crystal conversion from a thermodynamically stable phase.  During the preparation of our FeS samples, we found that maintaining a reducing and basic hydrothermal environment was crucial to observing superconductivity in FeS. The de-intercalation of potassium cations from K${_x}$Fe${_{2-y}}$S${_2}$ resulted in the shift of alternating planes of FeS along the $a$ direction of the unit cell to form the primitive layered FeS (Figure \ref{fgr_XRD}).  Note that Lei {\it et al.} had found K${_x}$Fe${_{2-y}}$S${_2}$ to be non-superconducting,\cite{Lei2011a} so our reductive de-intercalation technique tunes this spin glassy material into a superconductor.  

A similar structural transformation from a body-centered tetragonal structure to a primitive tetragonal structure has also been previously observed in the selenide analogue, K${_x}$Fe${_{2-y}}$Se${_2}$.\cite{Shoemaker2012} When exposed to air or moisture, oxidation of iron and formation of iron vacancies was suggested to be the driving force for the structural transition. After the structural change induced by oxidation in water, the superconducting K${_x}$Fe${_{2-y}}$Se${_2}$ became non-superconducting.\cite{Shoemaker2012} In contrast, our reductive de-intercalation was driven by preference of potassium cations to solvate into solution under strongly basic conditions, which consequently alters the non-superconducting K${_x}$Fe${_{2-y}}$S${_2}$, Figure S1 in Supplementary Materials (SM), into superconducting FeS. Also, the reducing environment in the autoclave maintained by the presence of Fe metal as a reagent prevented oxidation of Fe$^{2+}$ to Fe$^{3+}$ or the formation of iron vacancies.

A more drastic structural change could be possible under stronger oxidizing conditions. Neilson and McQueen\cite{Neilson2012} reported that KNi${_{2}}$Se${_2}$, a Ni analogue of the K${_x}$Fe${_{2-y}}$Se${_2}$, forms hexagonal NiAs-type, K${_{1-y}}$Fe${_{2-z}}$Se${_2}$, by oxidative de-intercalation of K${^+}$ by CuI${_2}$ in acetonitrile. This caused a complete structural reconstruction from edge-sharing layered NiSe${_4}$ tetrahedra to corner-sharing NiSe${_6}$ octahedra. Such a reconstruction was not seen in our de-intercalation reaction of K${_x}$Fe${_{2-y}}$S${_2}$ since we did not utilize strong oxidizing environment but rather maintained reducing conditions.  We similarly found this strategy in achieving the highest $T_c$'s for the (Li$_{1-x}$Fe$_x$OH)FeSe and (Li$_{1-x}$Fe$_x$OD)FeSe single crystals in their single crystal-to-single crystal conversion also utilizing K${_x}$Fe${_{2-y}}$Se${_2}$ as the template.\cite{Zhou2015}  A similar method was used for ion exchange in the single-crystal conversion of the selenide analogues K${_x}$Fe${_{2-y}}$Se${_2}$ to (Li$_x$Fe$_{1-x}$OH)FeSe,\cite{Dong2015} which demonstrates how powerful this technique is for exploring new layered iron chalcogenides.


\begin{figure}[t]
\centering
\includegraphics[width=0.9\columnwidth]{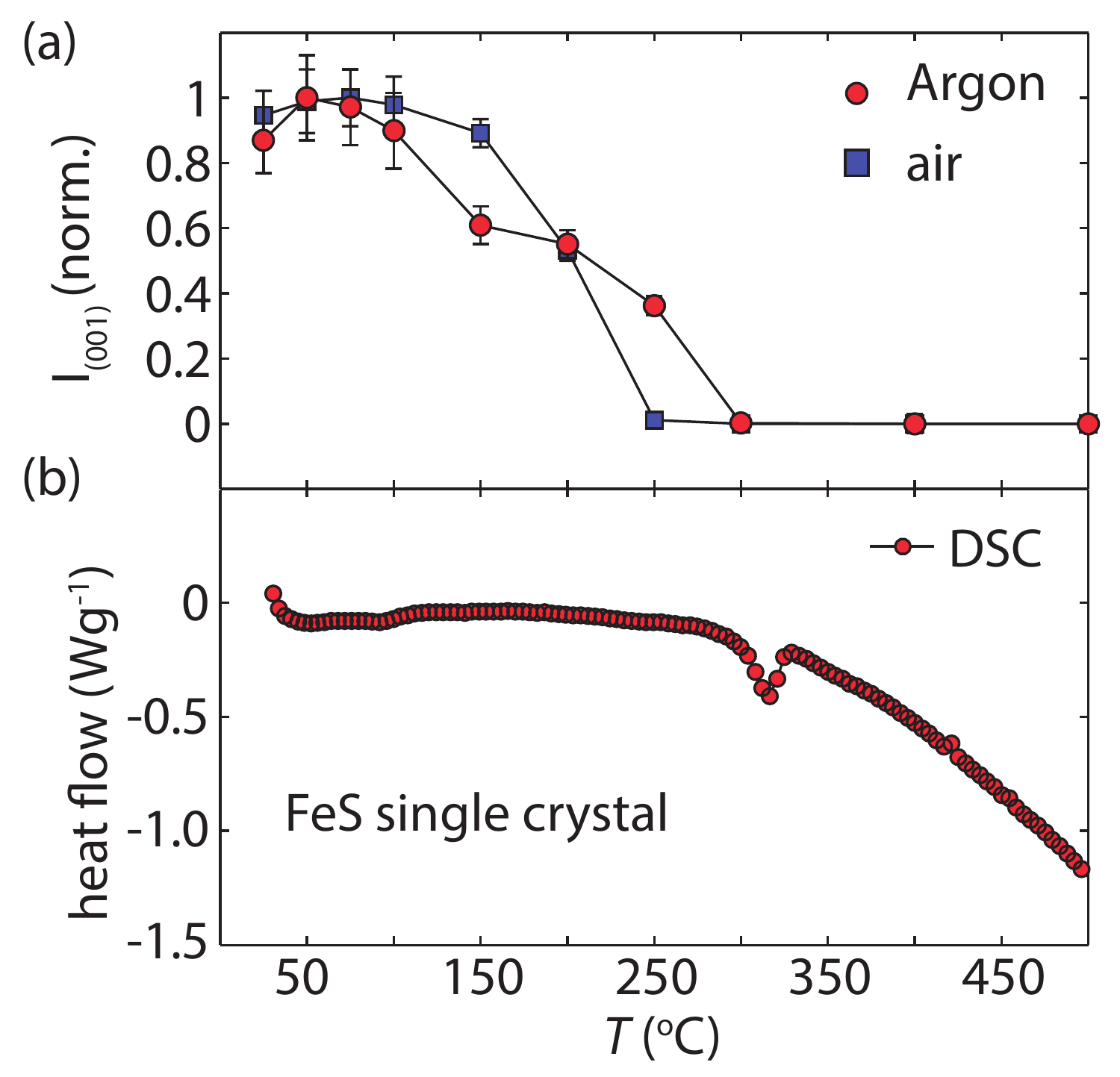}
\caption{ (a) Normalized integrated intensity of the (001) peak (top) from temperature dependent XRD. Under Argon (red curve), the loss of the (001) peak is gradual and is absent above 250 $^\circ$C. Under air (blue curve), the loss of (001) peak is more abrupt and the peak is absent above 200 $^\circ$C. (b) DSC results, plotted as heat flow as a function of temperature, for single crystal FeS. The sudden change in heat flow at 300 $^\circ$C is associated with an endothermic reaction.}
\label{fgr_TGA}
\end{figure}

\subsection{X-ray diffraction and crystal structure}

The XRD powder pattern of ground single crystals of K${_x}$Fe${_{2-y}}$S${_2}$, presented in Figure~\ref{fgr_XRD}a, shows pure crystalline product before the de-intercalation reactions. The pattern for K${_x}$Fe${_{2-y}}$S${_2}$ was fit with a body-centered tetragonal structural model with space group $I4/mmm$ and lattice parameters $a$ = 3.745(1) \AA ~and $c$ = 13.627(9) \AA\,(Table \ref{crystal_data}, Figure \ref{fgr_XRD}).  Full structural parameters from the fits are presented in Table \ref{crystal_data} and are in good agreement with those presented in an earlier study.\cite{Lei2011a}   Recently, Pachmayer \textit{et al.} found that FeS powders prepared by hydrothermal methods remain tetragonal down to low temperatures;\cite{Pachmayr2016} while the heavier congeners, FeSe\cite{McQueen2009, Margadonna2009, Boehmer2015} and FeTe,\cite{Rodriguez2011_a, Rodriguez2013} are known to have a crystallographic phase transitions.

After hydrothermal de-intercalation of potassium cations, the XRD pattern of the newly formed superconducting FeS crystals were fit to a primitive unit cell with space group ${P4/nmm}$ and lattice parameters ${a}$ = 3.6286(5) \AA ~and ${c}$ = 5.03440(9) \AA . These values were consistent with values previously reported for tetragonal FeS.\cite{Lennie1995, Lai2015, Denholme2014} Due to the layered nature of the samples, the XRD powder patterns for K${_x}$Fe${_{2-y}}$S${_2}$ and FeS were refined with preferred orientation along the [002] and [001] directions, respectively. Table \ref{crystal_data} presents the parameters of our structural refinements for ground single crystals of K${_x}$Fe${_{2-y}}$S${_2}$ and FeS as well as the powder samples of FeS prepared as a side reaction during the single-crystal-to-single-crystal conversion.  This powder consisted primarily of the product from the reaction of the iron powder in the presence of sodium sulfide and NaOH during the hydrothermal preparation (Figure S2 in SM).  For comparison, we have also prepared a powder sample of FeS through a modified method employed by Lai \textit{et al.},\cite{Lai2015} and the results from our diffraction measurements of a powder sample with $T_c = 4$ K are presented in the SM (Figures S3--S5 and Table S1 in SM).

\begin{table}[b!]
\caption{Structural parameters for ground single crystals of K${_x}$Fe${_{2-y}}$S${_2}$ and FeS along with FeS obtained through powder methods. Rietveld refinements with XRD data are of the room temperature structures.  In the FeS samples, we found full occupancy for the iron and sulfur sites. In the case of the K${_x}$Fe${_{2-y}}$S${_2}$ single crystals we found $x = 0.65(5)$ while $y$ was fixed to zero.  Relevant bond distances and angles are also included for each structural refinement.}
\resizebox{\columnwidth}{!}{
\begin{tabular}{l l l l l l }
\hline
\hline
\multicolumn{6}{c} {FeS (298 K, ground single crystal), $P4/nmm$, $R_{wp} = 3.042\%$ } \\
\hline
\multicolumn{6}{c} {$a = 3.6286$(5), $c = 5.03440$(9)} \\
atom		&      Site  		& 	x		&   y		&	z			& $U_{iso}$ (\AA$^2$)	\\
Fe1		&      2a		&	0  		&  	0 	 	&	0  			& 0.016(3) 					\\
S1			&      2c		&	0 		&   0.5	 	&	0.266(2) 	& 0.029(5)					\\
\\
S-Fe-S ($^\circ$)		& S-Fe-S ($^\circ$)		& Fe-S (\AA)		& Fe-Fe (\AA)	& anion height (\AA)	&{}	\\
108.1(2)					& 110.2(2)				& 2.275(5)		& 2.6040(5)		& 1.34(1) 				&{}	\\
\\
\hline
\multicolumn{6}{c} {FeS (298 K, powder preparation) , $P4/nmm$, $R_{wp} = 2.557\%$  } \\
\hline
\multicolumn{6}{c} {$a = 3.6841$(4), $c = 5.0334$(9)} \\
atom	&      Site 			& 	x  		& 	y	 	&	z			& $U_{iso}$ (\AA$^2$) 		\\
Fe1	&      2a	 		& 	0  		& 	0 		&	0 			& 0.034(3)					\\
S1		&      2c			& 	0  		& 	0.5		&  0.253(2) 	& 0.033(4)					\\
\\
S-Fe-S ($^\circ$)		& S-Fe-S ($^\circ$)		& Fe-S (\AA) 	& Fe-Fe (\AA) 	& anion height (\AA)	&{}	\\
110.7(4)					& 108.9(2)				& 2.239(5)		& 2.6051(4)		& 1.27(1)	 			&{}	\\
\\
\hline
\multicolumn{6}{c} {K${_x}$Fe${_{2-y}}$S${_2}$ (298 K, single crystal) , $I4/mmm$, $R_{wp} = 3.873\%$ } \\
\hline
\multicolumn{6}{c} {$a = 3.745$(1), $c = 13.627$(9)} \\
atom	&        Site  	&	x		&  	y	 	&  z		 	& $U_{iso}$ \AA$^2$ 	\\
K1		&        2a	 	&	0		& 	0	 	&  0		 	&  0.006(2)				\\
Fe1	&        4d	 	&   0		& 	0.5		&  0.25 		&  0.019(7)				\\
S1		&        4e	 	&   0  		& 	0		&  0.352(2) 	&  0.006(8)				\\
\\
S-Fe-S ($^\circ$)  		& S-Fe-S ($^\circ$) 		& Fe-S (\AA)		& Fe-Fe (\AA) 	& anion height (\AA) 	&{} \\
 110.8(3)	   				& 106.8(3)	 			& 2.33(2) 			& 2.6481(6)		& 1.39(3) 				&{} \\
\\
\hline
\hline
\end{tabular}}
\label{crystal_data}
\end{table}

\subsection{Thermal stability of FeS single crystals}

To test the thermal stability of our new FeS single crystals, samples were heated under inert Argon atmosphere in steps ranging from 25 $^{\circ}$C, 50 $^{\circ}$C, and 100 $^{\circ}$C. The 001 peak is visible up to 250 $^{\circ}$C (Figure S6 in SM), and its integrated intensity versus temperature under an Argon atmosphere is presented in Figure \ref{fgr_TGA}a along with a plot of the DSC. The decomposition of mackinawite FeS as determined by the integrated intensity of the (001) peak begin above 100 $^{\circ}$C and disappeared completely above 250 $^{\circ}$C. Due to the geometry of the XRD experiment, the (00$l$) reflections in the single crystal sample were observed while other reflections were not. Therefore, it is likely that if greigite were to form above $T$ = 100 $^\circ$C, it would not have been detected in our experiment. 

DSC measurements of FeS in Argon up to 600 $^\circ$C, shown in Figure \ref{fgr_TGA}b, give some clues on the thermal behavior during the decomposition of mackinawite. The dip in the heat flow around 300 $^{\circ}$C indicates an endothermic reaction that could be associated with the crystallization of a phase such as pyrrhotite not seen in our temperature dependent diffraction studies. The appearance of this transition in the DSC after the disappearance of the (001) reflection in the XRD, indicates that the two are related.  XRD analysis on the residue from the DSC experiment indicated formation of hexagonal pyrrhotite (Figure S7 in SM).   The higher than expected thermal stability of the mackinawite compared to past studies could be due to the single crystalline nature of our samples, which have larger surface areas and are therefore less reactive than a polycrystalline product with small particle sizes.

From their high-resolution X-ray diffraction study, Lennie \textit{{\it et al.}} reported that mackinawite begins to decompose to greigite (Fe$_3$S$_4$) above 100 $^\circ$C and that all FeS reflections disappear above $T$ = 200 $^\circ$C under a He atmosphere.\cite{Lennie1997} Above 260 $^\circ$C, greigite decomposes and hexagonal pyrrhotite begins to emerge.\cite{Lennie1997} 

Lennie \textit{{\it et al.}} also reported that mackinawite-FeS rapidly oxidizes under air.\cite{Lennie1995}. To test the air stability of our single crystals, we heated samples under ambient atmosphere in steps ranging from 25 $^{\circ}$C, 50 $^{\circ}$C, and 100 $^{\circ}$C. As presented in Fig \ref{fgr_TGA}b, the (001) peak is visible up to 200 $^{\circ}$C. As this level of air stability has not been reported for mackinawite before, it could imply that there may be some alkali metal incorporation that could passivate the surface and prevent oxidation of FeS. EDS mapping  on the surface of FeS single crystals shows up to 9\% total alkali (K and Na) on the surface of the FeS crystals (Figure S8 in SM). Due to the similarity of the $c$-parameter to those previously reported FeS, it is unlikely that large cations such as sodium or potassium intercalate between layers.


\section{Results: Physical properties}

\subsection{Magnetic susceptibility}

\begin{figure}[b!]
\centering
  \includegraphics[width=0.75\columnwidth]{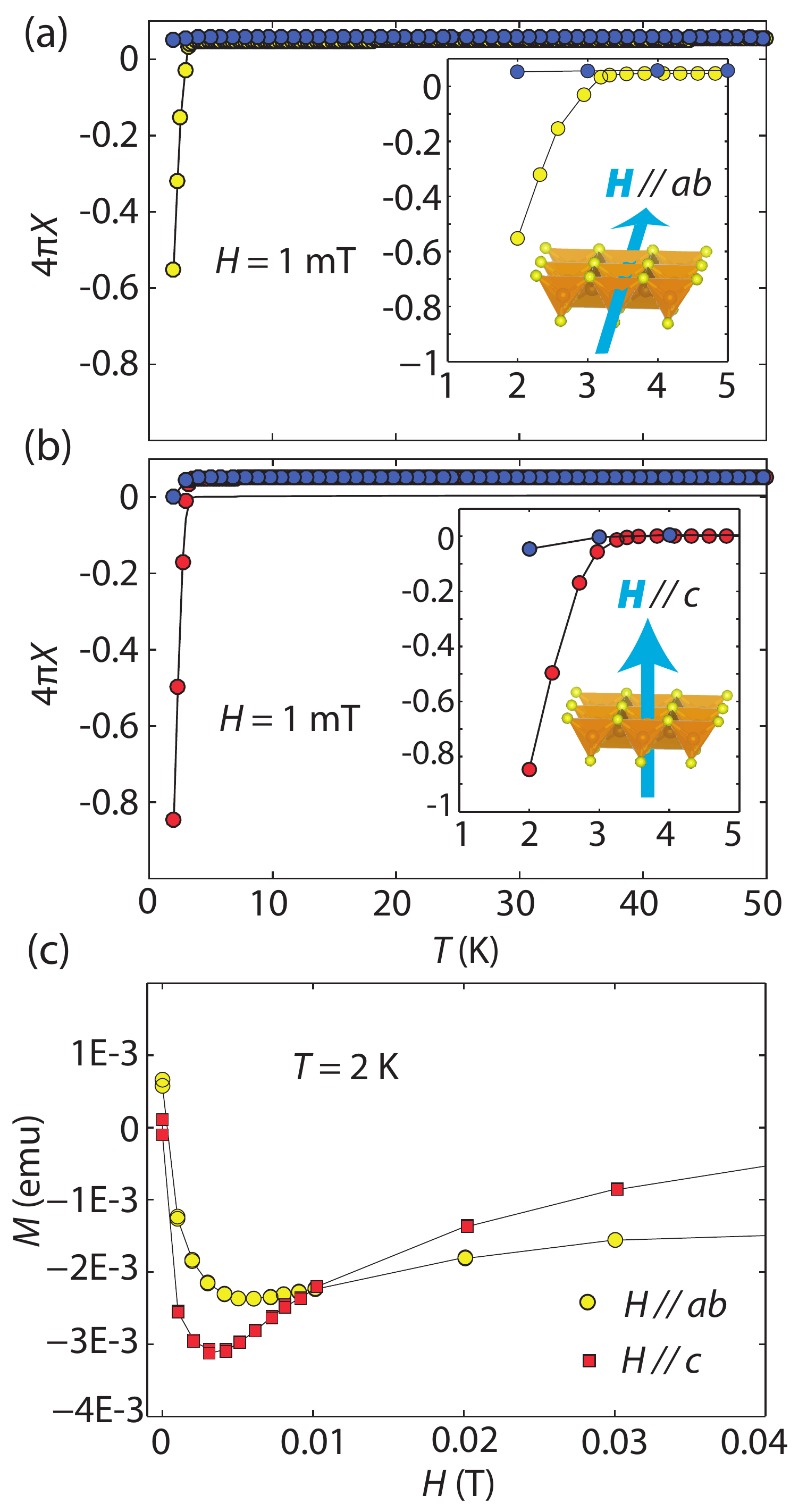}
  \caption{Magnetic susceptibility of an FeS single crystal. (a) Temperature-dependent volume susceptibility $4\pi\chi$ of an FeS crystal with a $H||{ab}$ shows Pauli paramagnetic behavior in the normal state and transitions to the superconducting at $T_c$ = 3.5~K. (b) Susceptibility for $H^{||c}$ with an increased diamagnetic response with a relative volume fraction increase of 30\% (c) Magnetization $M$ as a function of applied field at 2~K. The diamagnetic response weakens for fields greater than 4 mT ($H^{||ab}$) and 5 mT ($H^{||c}$).}
  \label{fgr_mag}
\end{figure}

The temperature-dependent FC and ZFC magnetic susceptibilities of FeS crystals measured in a constant field of 1~mT are presented in Figure~\ref{fgr_mag}, for fields applied both parallel and perpendicular to the crystallographic $c$-axis.   The volume susceptibility $4\pi\chi$ under ZFC conditions exhibits an onset superconducting transition at $T_c = 3.5$~K and a shielding fraction of $4\pi\chi \approx$60-90\% (without geometric factors taken into account). The significant superconducting volume fractions indicate that FeS is a bulk superconductor.  In both cases of the field orientation, the ZFC and FC curves  in the normal state above $T_c$ are largely temperature independent, indicative of Pauli paramagnetism and therefore metallicity in FeS.

Figure~\ref{fgr_mag}c presents magnetization ($M$) as a function of applied field ($H$) along two different directions for the applied field. The $M(H)$ isotherms indicate the values of the lower critical field $H_{c1}$ to be 4~mT and 5~mT at 1.8~K for $H||ab$ and $H||c$, respectively. One difference between our single crystal results and those of Lai {\it et al.} is the maximum critical temperature observed.  Lai {\it et al.} reported the superconducting powder samples of FeS to have a $T_c$ = 4.5 K,\cite{Lai2015} which is approximately 1 K greater than found for our single crystals. Magnetic susceptibility of our own prepared powder samples show $T_c^{onset}$ = 4 K (Figures S4 and S5 in SM).

\subsection{Heat capacity}

\begin{figure}[t]
\centering
  \includegraphics[width=0.75\columnwidth]{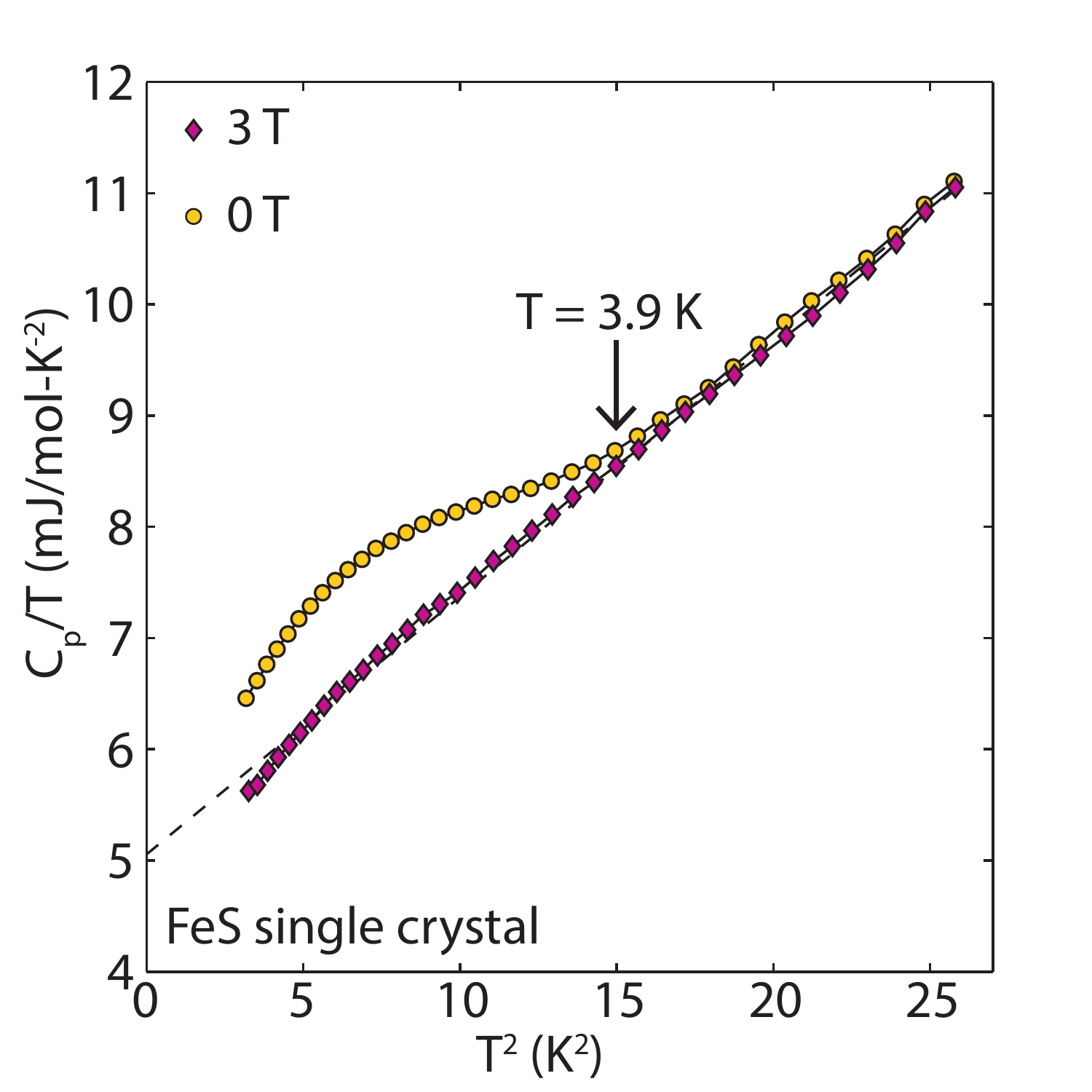}
  \caption{Low temperature specific heat of single crystal FeS for 0~T and 3~T applied magnetic fields. The arrow indicates the onset of a superconducting feature at $T$ = 3.9 K.}
  \label{fgr_HC}
\end{figure}

Heat capacity was measured on a large single crystal in both the superconducting (0 T) and normal (3 T) states. As shown in Figure \ref{fgr_HC}, a 3 T field is large enough to suppress the superconducting state in the crystal, making for a good comparison with the 0~T curve. 

In zero applied field, a clear signature of the superconducting transition develops at $T_c$=3.9 K, consistent with magnetic susceptibility and resistivity (below) measurements, confirming bulk superconductivity in single crystal FeS. 
Fitting the 3~T data to a standard electron and phonon contribution specific heat model,  $C = \gamma T + \beta T^3$, yields a normal state Sommerfield coefficient to be $\gamma$=5.1 mJ/mol-K$^2$ and phonon term $\beta$=0.23~mJ/mol-K$^4$, the latter corresponding to a Debye temperature $\Theta_D$= 257~K. Unlike reports for FeSe where the specific heat was fit to $C = \gamma T + \beta_3 T^3 + \beta_5 T^5$, \cite{McQueen2009}, for FeS a plot for $C/T$ vs $T^2$ is linear in the normal state. FeS does share some similarities with FeSe, however, as $\gamma$ was estimated to be 5.4(3) mJ/mol-K$^2$,\cite{McQueen2009}, which is within error to the value we found for $\gamma$ in FeS.

\subsection{Magnetoelectric transport}

\begin{figure}[t!] 
\centering
  \includegraphics[width=0.8\columnwidth]{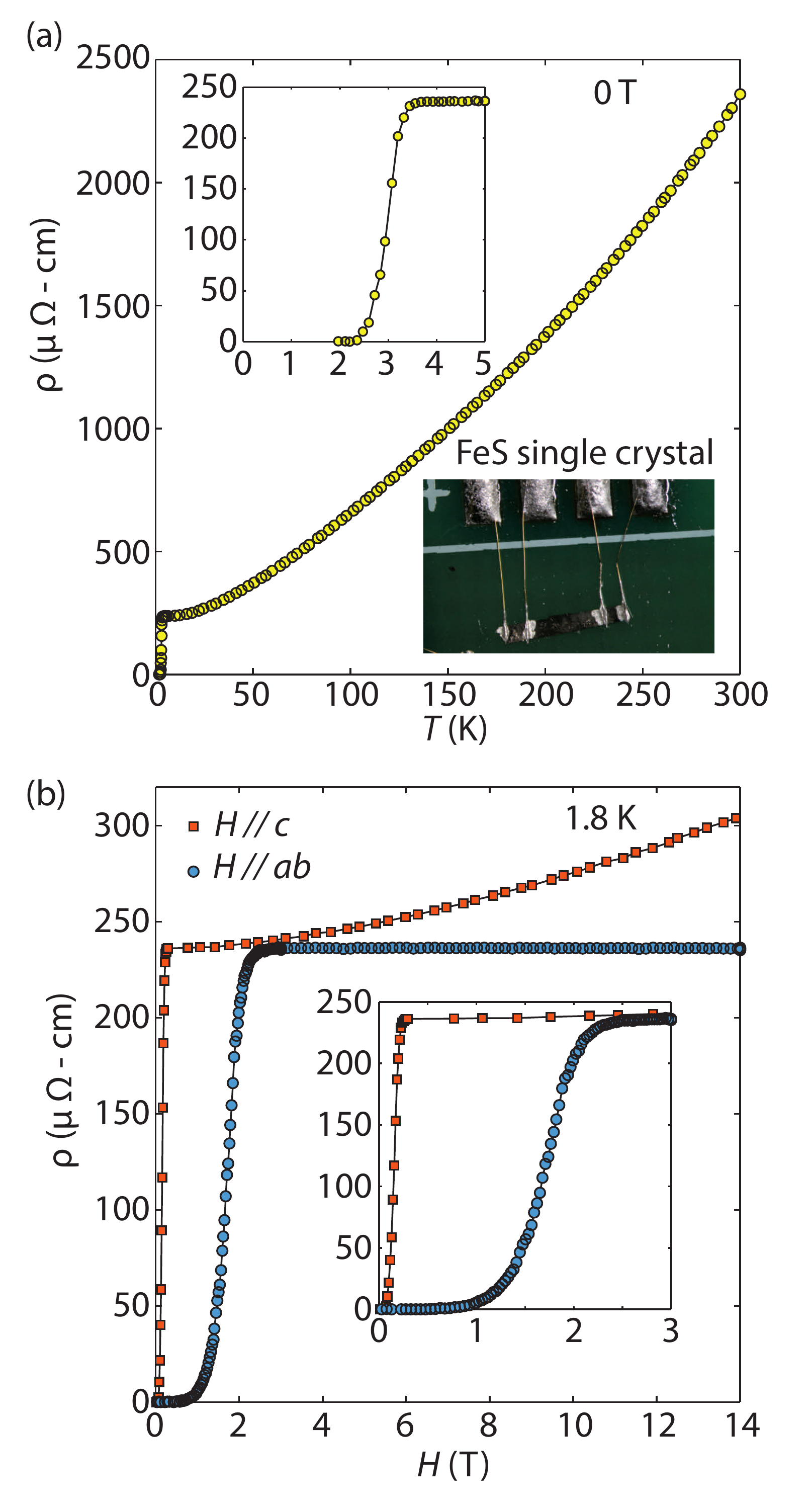}
  \caption{Electrical resistivity of single crystalline FeS. (a) Temperature-dependent resistivity with inset highlighting low temperature transition to the superconducting state at $T$ = 3.5 K. The geometry of the resistivity measurement for the single crystal also shown as inset.  (b) Resistivity as a function of applied magnetic field for both $H^{||ab}$ and $H^{||c}$ orientations (always transverse to current direction).}
  \label{fgr_res}
\end{figure}

Temperature dependent electrical resistivity of single-crystal FeS is presented in Figure~\ref{fgr_res}a. The resistivity exhibits metallic character down to the superconducting state with $T_c^{onset}$ = 3.5 K and $T_c^{zero}$ = 2.4 K. 
The residual resistivity of FeS was determined to be $\rho_0=240\, \mu\Omega\cdot$cm based on an average of the values measured for several samples (Figure S9 in SM), all of which exhibit a room temperature to residual resistivity ratio (RRR) of approximately 10, indicative of the high quality of our crystalline samples and the low uncertainty in geometric factors that may vary widely due to the micaceous nature of the crystals.

\begin{figure}[t!] 
\centering
  \includegraphics[width=0.8\columnwidth]{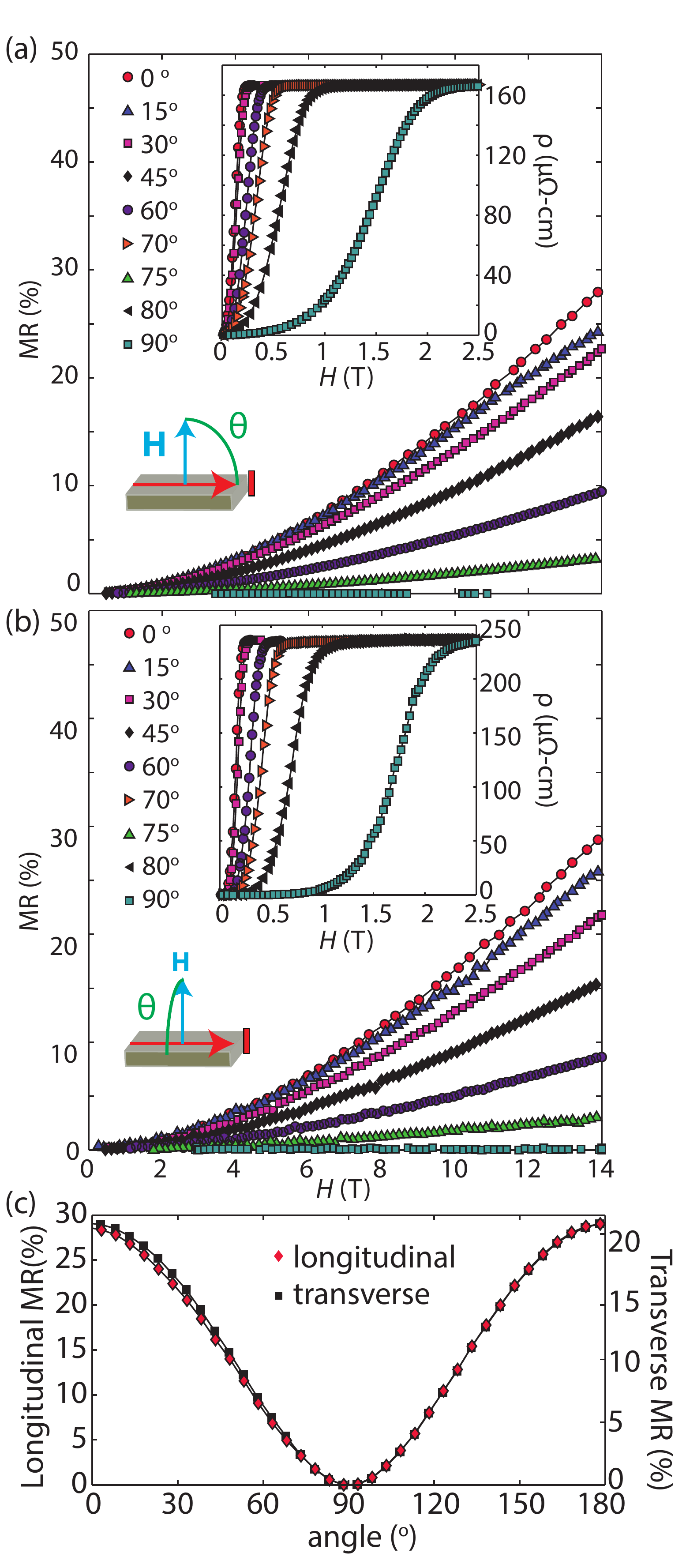}
  \caption{Constant temperature scans of magnetoresistance (MR) of FeS as a function of field angle $\theta$, defined as the deflection from $c$-axis direction. Angular dependence of (a) longitudinal ($H(90^{\circ})\parallel I$) and (b)  transverse ($H(\theta)\perp I$) MR taken at 1.8~K are presented. Insets in each figure display a zoom of the superconducting $H_{c2}$ transition. (c) Comparison of angular dependence of transverse and longitudinal MR at 1.8 K and 14 T.}
  \label{fgr_Rtheta}
\end{figure}

Figure~\ref{fgr_res}b presents the normalized magnetoresistance (MR) as a function of applied magnetic field at 1.8 K. As shown, a significant anisotropy appears in both the normal state high-field MR as well as the $H_{c2}$ transition, with the latter ranging from 0.16~T for $H\parallel c$ to 1.6~T for $H\parallel ab$.
The full angular dependence of these features are presented in Figure~\ref{fgr_Rtheta}. 
Panels (a) and (b) present the angular variation of MR for both longitudinal ($H(90^{\circ})\parallel I$) and transverse ($H(\theta)\perp I$) orientations, respectively. Indeed, as shown in Figure~\ref{fgr_Rtheta}c, the MR angular variation is well represented by a cosine-like dependence for both longitudinal and transverse orientation angles. 

\begin{figure}[t!]
\centering
  \includegraphics[width=0.95\columnwidth]{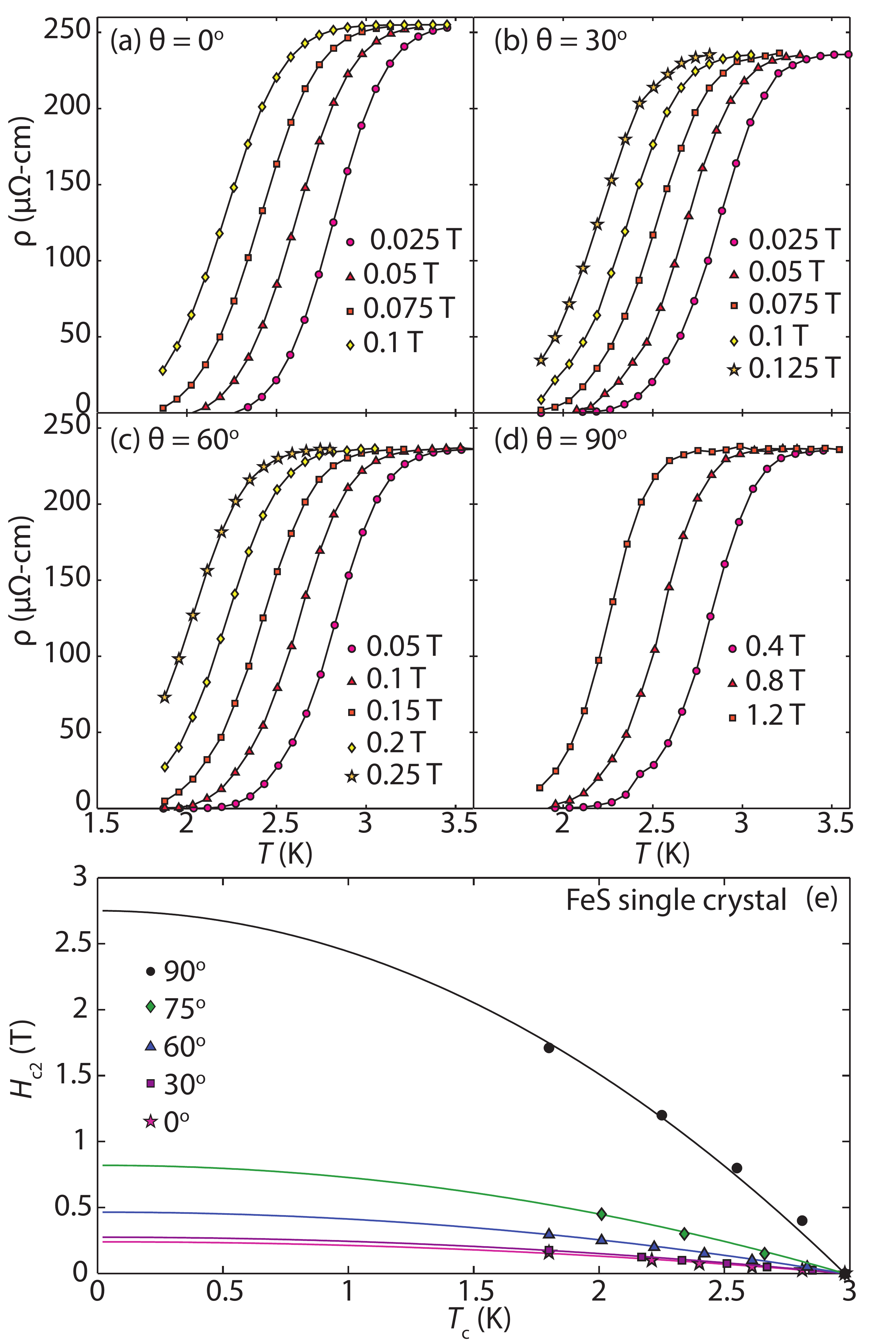}
  \caption{Superconducting transition of FeS single crystal as a function of magnetic field applied along different angles $\theta$ with respect to the crystallographic $c$-axis (transverse to current direction). Panel e) presents the compiled $H_{c2}(\theta)$ temperature dependences for all angles measured. $T_c$ values were determined by the resistance transition midpoint. Solid lines represent the WHH orbital pair-breaking expectation for $H_{c2}(T)$ in each case (see text for details).}
  \label{fgr_Hc2}
\end{figure}

A very large anisotropy is also evident in the upper critical field $H_{c2}$ as the field angle is rotated away from the $c$-axis. In both longitudinal and transverse orientations, $H_{c2}$ is observed to diminish strongly as the field rotates toward the basal plane, as shown in the insets of Figure~\ref{fgr_Rtheta}a-b. Taking the two extremes, one can define an $H_{c2}$ anisotropy $\Gamma \equiv H_{c2}^{||ab}/ H_{c2}^{||c}$, which is a value of 10 at 1.8~K. A more complete evaluation of the full  $H_{c2}(T)$ dependence allows for an extrapolation of $\Gamma$ to zero temperature. As shown in Figure~\ref{fgr_Hc2}a-d, extracting the $H_{c2}(T)$ values from the resistive transitions at several angles (all transverse to current direction, with $T_c$ values chosen at the 50\% resistance midpoint) leads to a full $H_{c2}(T)$ plot given in Figure~\ref{fgr_Hc2}e. For all field directions, \textit{H$_{c2}$(0)} was estimated using the Werthamer-Helfand-Hohenberg (WHH) formula ($H_{c2} = 0.69[-(dH_{c2}/dT)]|_{T_c}T_c$).\cite{Werthamer1966} Fitting results give $H_{c2}^{||ab}(0)$ = 2.75~T and $H_{c2}^{||c}(0) = 0.275$~T, yielding nearly the same anisotropy value $\Gamma(0)$=10 as for 1.8~K.
The coherence lengths calculated from the estimated $H_{c2}(0)$ values ($\xi = \sqrt{\Phi_\circ/({2\pi H_{c2}})}$ where $\Phi$ is the flux quantum) are calculated to be $\xi_{H||ab}$ = 343~\AA ~and $\xi_{H||c}$= 104~\AA.

\begin{figure}[t!] 
\centering
  \includegraphics[width=0.8\columnwidth]{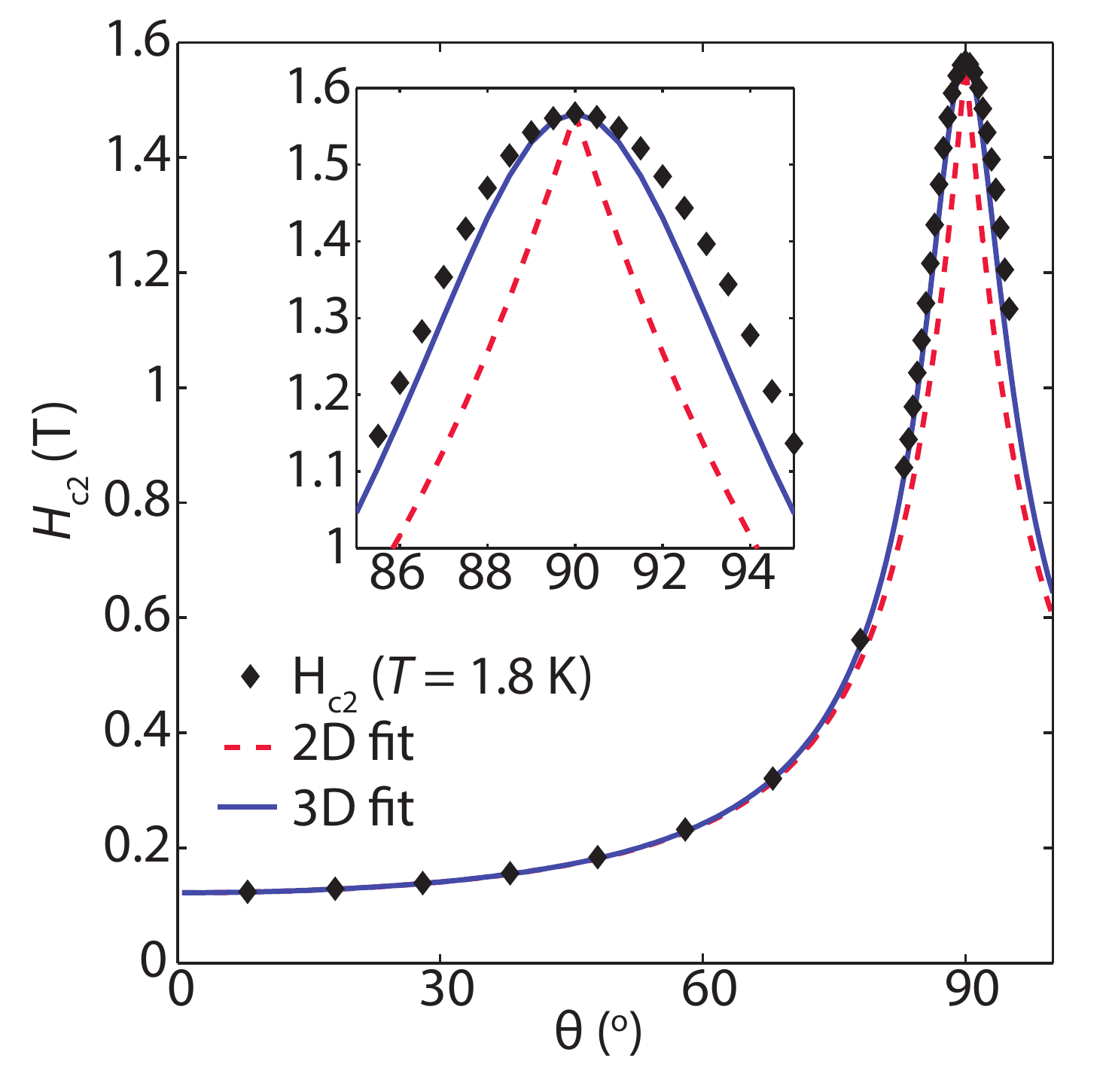}
  \caption{Angular dependence of the superconducting upper critical field $H_{c2}$ at 1.8~K. Black diamonds represented measured transition fields (defined as the resistive transition midpoint), and lines represent fits to the theoretical expectation for the angular dependence: solid line represents the Ginzburg Landau expectation for a 3D system with anisotropic effective mass, and dashed line represents the Tinkham's model expectation for 2D superconductors. Inset displays zoomed data near $90^{\circ}$ ($H\parallel ab$). All data were collected with magnetic field direction always transverse to the current direction.}
  \label{fgr_Hc2theta}
\end{figure}

These large changes in $H_{c2}$ with field angle and the concomitant coherence length anisotropy are in line with the strong anisotropy observed in the normal state MR as discussed above. To determine whether the large $H_{c2}$ anisotropy is indicative of a truly two-dimensional and not a strongly anisotropic three-dimensional superconducting system, we performed detailed measurements of the angular dependence of  $H_{c2}$ at 1.8 K. Figure~\ref{fgr_Hc2theta} presents the angle dependence of $H_{c2}$(1.8~K) as determined from midpoints of field sweep resistive transitions. (Using different criterion to define $H_{c2}$ results in slight variation in absolute anisotropy, but the shape of the $H_{c2}(\theta)$ curve remains constant). 
The shape of the $H_{c2}(\theta)$ curve, especially near the $H\parallel ab$ ($\theta=90^{\circ}$) orientation, is indicative of the true dimensionality of the superconductor with respect to the coherence length. 
Tinkham's model for thin-film superconductors incorporates the effect of reduced dimensionality, \cite{Tinkham1963} yielding an angular dependence given by
\begin{equation}
\left |\frac{H_{c2}(\theta){\rm sin}\theta}{H_{c2}^{\perp}}\right| + \left(\frac{H_{c2}(\theta){\rm cos}\theta}{H_{c2}^{\parallel}}\right)^2=1,
\end{equation}
whereas Ginzburg Landau (GL) theory \cite{Ketterson1999} can be used to determine the effect of an anisotropic effective mass $m^*$ on the angular dependence as
\begin{equation}
\left(\frac{H_{c2}(\theta){\rm sin}\theta}{H_{c2}^{\perp}}\right)^2+\left(\frac{H_{c2}(\theta){\rm cos}\theta}{H_{c2}^{\parallel}}\right)^2=1.
\end{equation}
As shown in the inset of Figure~\ref{fgr_Hc2theta}, the $H_{c2}(\theta)$ data is much better represented by the anisotropic GL theory, suggesting a highly anisotropic 3D environment for the superconductivity in FeS. This can be quantified by using the calculated anisotropy for this sample $\Gamma\simeq 12.8$ to extract the effective mass ratio $m^*_{\parallel}/m^*_{\perp}$=$\Gamma^2$=164. This is believed to be the largest upper critical field anisotropy observed in any Fe based superconductor reported so far.

\section{Discussion}

\subsection{Strongly anisotropic electronic properties}

The previous report for powder samples of FeS found $H_{c2}(0)$ to be 0.4 T,\cite{Lai2015} which is much lower than that of FeSe and other iron-based superconductors. $H_{c2}$ for FeSe has been reported to be 16.3 T in powder samples.\cite{Hsu2008}  This difference between the upper critical fields in FeSe and FeS has significant effects on their coherence lengths as well. Coherence lengths calculated from $H_{c2}(0)$ for FeS powders\cite{Lai2015} and FeSe powders\cite{Hsu2008} are 287 \AA \, and 45.0 \AA, respectively.  We confirm Lai's report of a much lower $H_{c2}(0)$ and higher coherence length in FeS compared to other iron-based superconductors, but also demonstrate that these properties are highly anisotropic.  

As important as the comparatively smaller critical fields in FeS, the anisotropy also appears to be much larger in this system.  We find an anisotropy ratio of $\Gamma \sim 10$, and to our knowledge this is the largest reported $\Gamma$ yet for an iron-based superconductor.   For FeTe$_{1-y}$S$_y$ single crystals, the field dependence on $T_c$ is mostly isotropic with a reported $\Gamma= H_{c2}^{||ab} / H_{c2}^{||c}$ = 18 T / 19 T = 0.95.\cite{Hu2009}.  Recent studies on Fe(Se$_{1-x}$S$_{x}$) single crystals has shown sulfur to increase $T_c$ from 8.5 K for $x$ = 0 to 10.7 K for $x$ = 0.11, and the anisotropy is also more pronounced in crystals with higher sulfur content as $\Gamma = H_{c2}^{||ab} / H_{c2}^{||c} = 2$ for $x = 0$ and 3.5 for $x = 0.11$.\cite{Abdel-Hafiez2015} 


Surprisingly, in our studies of angular dependence of MR, both longitudinal and transverse rotation studies show a diminishment of MR as the field is rotated toward the crystallographic basal plane, irrespective of whether the field direction is rotated parallel or perpendicular to the current direction (Figure \ref{fgr_Rtheta}a,b). This is consistent with either a projection-like orbital MR of a very thin specimen ({\it i.e.}, with a large MR when $H$ is perpendicular to the plane where orbital motion is allowed and zero MR when orbital motion of charge carriers is prohibited by geometric confinement), or with a very strong electronic anisotropy as found in other materials with reduced electronic dimensionality.

Given the micaceous nature of FeS single crystals, the anisotropic behavior of the MR may arise due to a microscopic physical separation of crystalline layers resulting in effectively two-dimensional layers that would act much as in a thin film.    Such a description of our sample's behavior would imply that it contains a slab thickness that is less than the characteristic magnetic length scale. 
Our studies of $H_{c2}$ anisotropy and its angular variation (Figure \ref{fgr_Hc2theta}) suggest that the measured superconducting state of FeS is in fact inhabiting a three-dimensional environment with strong anisotropy, given the lack of a cusp in $H_{c2}(\theta)$ near the $90^{\circ}$ field alignment (Figure \ref{fgr_Hc2theta}).  The result for our case is in good agreement with GL theory.
Therefore, the appropriate length scale to consider is the superconducting coherence length which is 104~\AA\ for $\xi_{H||ab}$. In other words, our single-crystal samples must entail crystalline slabs of at least 104~\AA\, thickness in order to exhibit the GL-type behavior of $H_{c2}$ that follows from Eq. 2. 
An estimate of the mean free path of quasiparticles \cite{Millis1988} yields $l_{\rm mfp}\approx 30$~\AA, which is much smaller than 104~\AA, suggesting the scattering length is at least much smaller than the known slab thickness. At the very least, the fact that the effective thickness must be at least $\sim$20 unit cells suggests quasiparticles are not artificially confined, and that the the observed two-dimensional behavior in MR may be intrinsic to the electronic structure.

\subsection{True ground and normal state properties of FeS}

The tetragonal FeS system was originally predicted to be semiconductor in nature by Bertaut \textit{{\it et al.}}\cite{Bertaut1965} This claim was recently supported by resistivity measurements performed by Denholme \textit{{\it et al.}}\cite{Denholme2014}, which showed that their samples were non-superconducting with ferrimagnetic-like behavior. Similarly, samples prepared by Sines \textit{{\it et al.}}\cite{Sines2012} were also exhibited semiconducting and ferrimagnetic behavior. Contrary to experimental evidence published before the work of Lai \textit{et al.},\cite{Lai2015} several other groups had predicted tetragonal FeS to be metallic.\cite{Vaughan1971, Subedi2008, Kwon2011, Devey2008, Brgoch2012} Vaughan and Ridout\cite{Vaughan1971} proposed that the bonding in the tetragonal FeS was metallic in nature due to delocalized \textit{d} electrons in iron sublattice. Recent density functional theory (DFT) calculations also supported metallicity, in tetragonal FeS.\cite{Subedi2008, Kwon2011, Devey2008}

Geochemists studying mackinawite have suggested that the ferrimagnetic-like behavior from earlier magnetization data might have risen from the well-known thiospinel ferrimagnetic impurity, Fe$_3$S$_4$, considering the ease of conversion of mackinawite FeS to Fe$_3$S$_4$.\cite{Lennie1995a, Malasics2012} Several of our powder FeS samples prepared through the synthesis detailed by Lennie \textit{et al.}\cite{Lennie1995} form with an Fe$_3$S$_4$ impurity as revealed by combined magnetization measurements and neutron powder diffraction (Figures S10-12 in SM). Even Denholme \textit{ et al.} acknowledged that the semiconductor behavior of FeS could be attributed to the surface oxide layers of FeS, as suggested by Bertaut \textit{et al.} \cite{Denholme2014, Bertaut1965} Indeed, similar oxidation has been observed in the FeSe system, as Greenfield \textit{et al.}\cite{Greenfield2015} reported that amorphous surface oxide layers of FeSe particles suppressed the superconductivity in FeSe.  Our single crystal results definitively support a metallicity in the normal state properties and superconductivity in the ground state.

\subsection{Structural trends concerning $T_c$}

Compared to tetragonal FeSe, mackinawite FeS contains more regular tetrahedral $Ch$--Fe--$Ch$ bond angles where $Ch=$ chalcogenide. In FeSe, the Se--Fe--Se out-of-plane bond angle is 112.32(6)$^{\circ}$ and the Se--Fe--Se in-plane bond angle is 103.91(7)$^{\circ}$.\cite{McQueen2009} The respective bond angles for our FeS powder and single crystal samples were calculated to be close to 108.1(3)$^{\circ}$ and 110.2(2)$^{\circ}$ (Table \ref{crystal_data}). Several studies have suggested that higher $T_c$ could be achieved from more regular bond angles,\cite{Denholme2014} as is with iron pnictide superconductors.\cite{Lee2012, Mizuguchi2010} However, this structural parameter does not seem to be as important an indicator in the iron chalcogenides since FeSe exhibits a higher $T_c$ (8 K) than FeS ($T_c$ = 4 K) even though it is comprised of more distorted tetrahedra. This suggests that structural factors controlling $T_c$ in iron pnictides may not be identical to those of the iron chalcogenides.

\begin{figure}[t!] 
\centering
  \includegraphics[width=0.8\columnwidth]{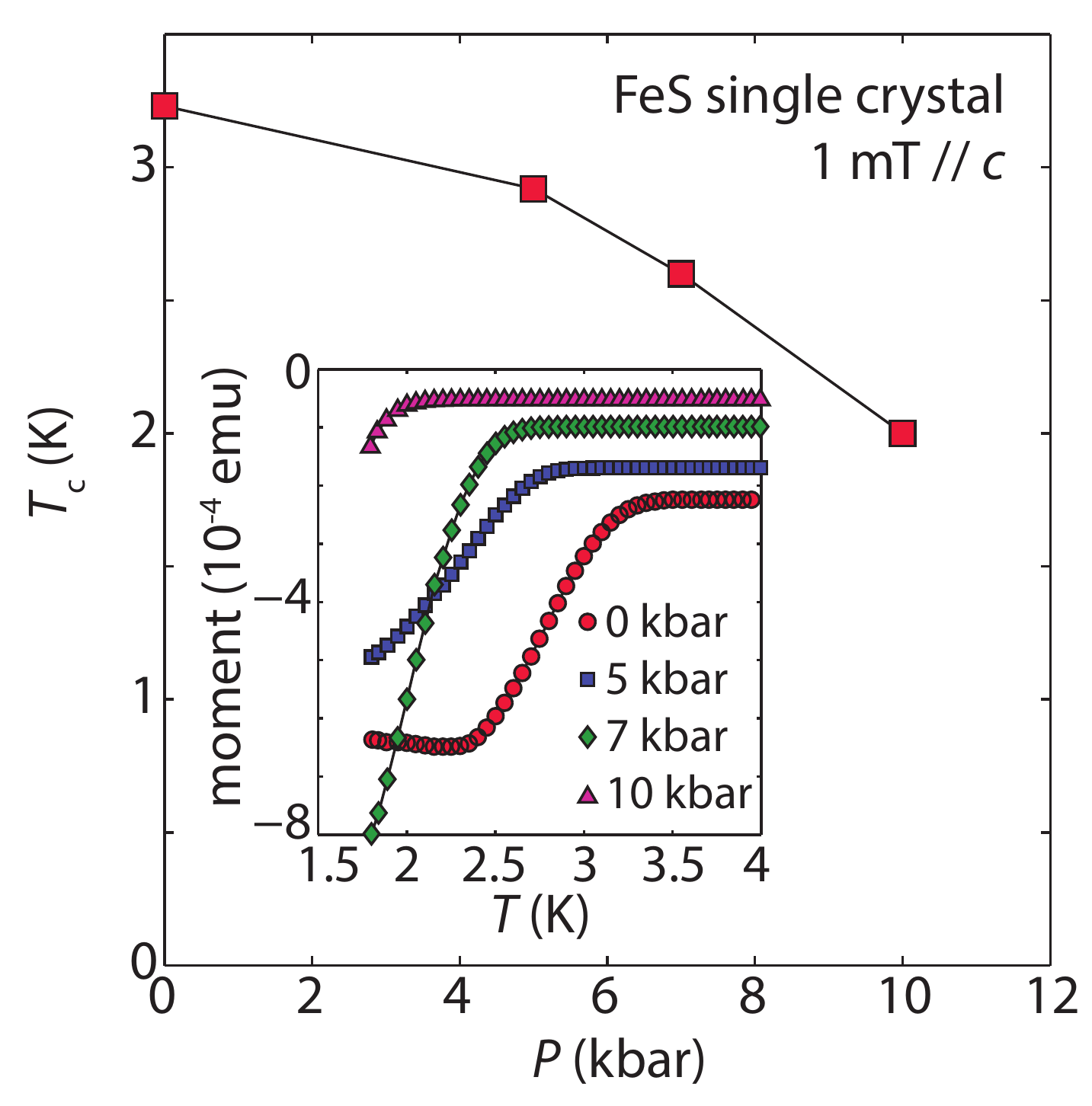}
  \caption{Applied pressure dependence of the superconducting transition temperature 
  of an FeS crystal as extracted from magnetic susceptibility measurements performed using a BeCu piston-cylinder clamp cell. Inset displays the measured susceptibility data (presented with a vertical offset for clarity). Pressure values are determined at room temperature.}
  \label{fgr_pressure}
\end{figure}

Anion height has also been implicated as a reliable predictor for $T_c$ in iron-based superconductors.\cite{Mizuguchi2010} For iron pnictides, $T_c$ increases with increasing anion height as FeP-based superconductors have lower anion height and lower $T_c$ than FeAs-based superconductors. However, $T_c$ begins to drop off for anion heights greater than 1.38 \AA, which suggests there is an optimal anion height for maximizing $T_c$. For FeSe with $T_c$ = 8 K, the Se height is 1.45 \AA, and upon application of physical pressure, the Se height decreases to 1.425 \AA, which leads to an increase in $T_c$ up to 37 K (8 GPa).\cite{Imai2009, Mizuguchi2010} For larger anions, i.e. FeTe, the anion height is larger than that of FeSe and while FeTe is not superconducting at ambient pressure isovalent anionic substitution as in FeTe$_{0.8}$S$_{0.2}$ induces superconductivity (anion height = 1.75 \AA, $T_c$ = 10 K).\cite{Mizuguchi2009, Zajdel2010} From this anion height principle, we should expect the smaller anionic radius of sulfide to lead to a larger $T_c$. However, the anion height in FeS was found in the range from 1.27(1) to 1.34(1) \AA \,(Table \ref{crystal_data}), which is below the optimal height of 1.38 \AA.  This result for FeS could therefore explain why the $T_c$ is  remains low and between 3.5 and 5 K despite having more regular tetrahedra than FeSe or FeTe.

As a preliminary study on modifying the anion height in FeS to affect $T_c$, we have performed magnetization measurements as a function applied pressure. As shown in Figure~\ref{fgr_pressure}, measurements of magnetic susceptibility in a clamp-cell setup show that the transition temperature decreases with increasing pressure, at least up to 10~kbar. While it is known that $T_c$ in the related superconductor FeSe undergoes a dramatic enhancement under pressure, the increase in $T_c$  for FeSe occurs at much higher pressures than currently reached in the present experiment for FeS (on the order of 10~GPa). Further work to study the relation between $T_c(P)$ and the crystallographic parameters as a function of applied pressure will shed more light on the relation between structure and superconductivity in FeS.

\section{Conclusions}

In conclusion, we have synthesized superconducting single crystals of FeS and characterized their thermal, magnetic, and electrical properties. The synthesis of FeS single crystals was accomplished through the novel method of reductive de-intercalation of K${_x}$Fe${_{2-y}}$S${_2}$ single crystals under hydrothermal conditions. The FeS crystals are stable up to 250 $^{\circ}$C in argon and 200 $^{\circ}$C in air.  
At 4~K the FeS crystals transition from a metallic, Pauli paramagnetic state to the superconducting state. In both the normal state and superconducting states, we observe a large anisotropy in the properties of FeS.  The upper critical field expresses a large anisotropy with a $ \Gamma = H_{c2}^{||ab}(0) / H_{c2}^{||c}(0)=(2.75~T)/(0.275~T)=10$, the largest reported for any iron-based superconductor thus far.  Magnetoresistance measurements for the normal state performed as a function of applied field angle reveal a remarkable two-dimensional behavior in FeS.  Overall, the physical property results indicate that the Fermi surface of FeS may be highly two-dimensional, and perhaps even more so than other closely-related iron-based superconductors.  Since the metastable system, mackinawite-type FeS, is now confirmed as a superconductor and not a magnetic semiconductor, this system could be a template for the preparation of new sulfide-based superconductors that exhibit strong anisotropic behavior.

\section{Acknowledgements}

Research at the University of Maryland was supported by the NSF Career DMR-1455118, AFOSR Grant No. FA9550-14-10332, and the Gordon and Betty Moore Foundation Grant No. GBMF4419. We acknowledge the support of the National Institute of Standards and Technology, U. S. Department of Commerce, in providing the neutron research facilities used in this work.



\begin{mcitethebibliography}{56}
\providecommand*{\natexlab}[1]{#1}
\providecommand*{\mciteSetBstSublistMode}[1]{}
\providecommand*{\mciteSetBstMaxWidthForm}[2]{}
\providecommand*{\mciteBstWouldAddEndPuncttrue}
  {\def\EndOfBibitem{\unskip.}}
\providecommand*{\mciteBstWouldAddEndPunctfalse}
  {\let\EndOfBibitem\relax}
\providecommand*{\mciteSetBstMidEndSepPunct}[3]{}
\providecommand*{\mciteSetBstSublistLabelBeginEnd}[3]{}
\providecommand*{\EndOfBibitem}{}
\mciteSetBstSublistMode{f}
\mciteSetBstMaxWidthForm{subitem}
{(\emph{\alph{mcitesubitemcount}})}
\mciteSetBstSublistLabelBeginEnd{\mcitemaxwidthsubitemform\space}
{\relax}{\relax}

\bibitem[Paglione and Greene(2010)]{Paglione2010}
J.~Paglione and R.~L. Greene, \emph{Nat. Phys.}, 2010, \textbf{6},
  645--658\relax
\mciteBstWouldAddEndPuncttrue
\mciteSetBstMidEndSepPunct{\mcitedefaultmidpunct}
{\mcitedefaultendpunct}{\mcitedefaultseppunct}\relax
\EndOfBibitem
\bibitem[Johnston(2010)]{Johnston2010}
D.~C. Johnston, \emph{Adv. Phys.}, 2010, \textbf{59}, 803--1061\relax
\mciteBstWouldAddEndPuncttrue
\mciteSetBstMidEndSepPunct{\mcitedefaultmidpunct}
{\mcitedefaultendpunct}{\mcitedefaultseppunct}\relax
\EndOfBibitem
\bibitem[Ivanovskii(2011)]{Ivanovskii2011}
A.~Ivanovskii, \emph{Physica C}, 2011, \textbf{471}, 409--427\relax
\mciteBstWouldAddEndPuncttrue
\mciteSetBstMidEndSepPunct{\mcitedefaultmidpunct}
{\mcitedefaultendpunct}{\mcitedefaultseppunct}\relax
\EndOfBibitem
\bibitem[Takahashi \emph{et~al.}(2015)Takahashi, Sugimoto, Nambu, Yamauchi,
  Hirata, Kawakami, Avdeev, Matsubayashi, Du, Kawashima, Soeda, Nakano,
  Uwatoko, Ueda, Sato, and Ohgushi]{Takahashi2015}
H.~Takahashi, A.~Sugimoto, Y.~Nambu, T.~Yamauchi, Y.~Hirata, T.~Kawakami,
  M.~Avdeev, K.~Matsubayashi, F.~Du, C.~Kawashima, H.~Soeda, S.~Nakano,
  Y.~Uwatoko, Y.~Ueda, T.~J. Sato and K.~Ohgushi, \emph{Nat. Mater.}, 2015,
  \textbf{14}, 1008--1012\relax
\mciteBstWouldAddEndPuncttrue
\mciteSetBstMidEndSepPunct{\mcitedefaultmidpunct}
{\mcitedefaultendpunct}{\mcitedefaultseppunct}\relax
\EndOfBibitem
\bibitem[Drozdov \emph{et~al.}(2015)Drozdov, Eremets, Troyan, Ksenofontov, and
  Shylin]{Drozdov2015}
A.~P. Drozdov, M.~I. Eremets, I.~A. Troyan, V.~Ksenofontov and S.~I. Shylin,
  \emph{Nature}, 2015, \textbf{525}, 73--76\relax
\mciteBstWouldAddEndPuncttrue
\mciteSetBstMidEndSepPunct{\mcitedefaultmidpunct}
{\mcitedefaultendpunct}{\mcitedefaultseppunct}\relax
\EndOfBibitem
\bibitem[Lai \emph{et~al.}(2015)Lai, Zhang, Wang, Wang, Zhang, Lin, and
  Huang]{Lai2015}
X.~Lai, H.~Zhang, Y.~Wang, X.~Wang, X.~Zhang, J.~Lin and F.~Huang, \emph{J.
  Amer. Chem. Soc.}, 2015, \textbf{137}, 10148--10151\relax
\mciteBstWouldAddEndPuncttrue
\mciteSetBstMidEndSepPunct{\mcitedefaultmidpunct}
{\mcitedefaultendpunct}{\mcitedefaultseppunct}\relax
\EndOfBibitem
\bibitem[Kouvo \emph{et~al.}(1963)Kouvo, Long, and Vuorelainen]{Kouvo1963}
O.~Kouvo, J.~Long and Y.~Vuorelainen, \emph{Am. Mineral.}, 1963, \textbf{48},
  511\relax
\mciteBstWouldAddEndPuncttrue
\mciteSetBstMidEndSepPunct{\mcitedefaultmidpunct}
{\mcitedefaultendpunct}{\mcitedefaultseppunct}\relax
\EndOfBibitem
\bibitem[Bertaut \emph{et~al.}(1965)Bertaut, Burlet, and Chappert]{Bertaut1965}
E.~Bertaut, P.~Burlet and J.~Chappert, \emph{Solid State Commun.}, 1965,
  \textbf{3}, 335 -- 338\relax
\mciteBstWouldAddEndPuncttrue
\mciteSetBstMidEndSepPunct{\mcitedefaultmidpunct}
{\mcitedefaultendpunct}{\mcitedefaultseppunct}\relax
\EndOfBibitem
\bibitem[Evans~Jr \emph{et~al.}(1964)Evans~Jr, Milton, Chao, Adler, Mead,
  Ingram, and Berner]{EvansJr1964}
H.~T. Evans~Jr, C.~Milton, E.~Chao, I.~Adler, C.~Mead, B.~Ingram and R.~A.
  Berner, \emph{US Geol. Survey Prof. Paper}, 1964, \textbf{475}, 1312 --
  1318\relax
\mciteBstWouldAddEndPuncttrue
\mciteSetBstMidEndSepPunct{\mcitedefaultmidpunct}
{\mcitedefaultendpunct}{\mcitedefaultseppunct}\relax
\EndOfBibitem
\bibitem[Lennie \emph{et~al.}(1995)Lennie, Redfern, Schofield, and
  Vaughan]{Lennie1995}
A.~Lennie, S.~Redfern, P.~Schofield and D.~Vaughan, \emph{Mineral. Mag.}, 1995,
  \textbf{59}, 677--684\relax
\mciteBstWouldAddEndPuncttrue
\mciteSetBstMidEndSepPunct{\mcitedefaultmidpunct}
{\mcitedefaultendpunct}{\mcitedefaultseppunct}\relax
\EndOfBibitem
\bibitem[Rickard and Luther(2007)]{Rickard2007}
D.~Rickard and G.~W. Luther, \emph{Chem. Rev.}, 2007, \textbf{107},
  514--562\relax
\mciteBstWouldAddEndPuncttrue
\mciteSetBstMidEndSepPunct{\mcitedefaultmidpunct}
{\mcitedefaultendpunct}{\mcitedefaultseppunct}\relax
\EndOfBibitem
\bibitem[Denholme \emph{et~al.}(2014)Denholme, Demura, Okazaki, Hara, Deguchi,
  Fujioka, Ozaki, Yamaguchi, Takeya, and Takano]{Denholme2014}
S.~Denholme, S.~Demura, H.~Okazaki, H.~Hara, K.~Deguchi, M.~Fujioka, T.~Ozaki,
  T.~Yamaguchi, H.~Takeya and Y.~Takano, \emph{Mater. Chem. Phys.}, 2014,
  \textbf{147}, 50 -- 56\relax
\mciteBstWouldAddEndPuncttrue
\mciteSetBstMidEndSepPunct{\mcitedefaultmidpunct}
{\mcitedefaultendpunct}{\mcitedefaultseppunct}\relax
\EndOfBibitem
\bibitem[Sines \emph{et~al.}(2012)Sines, Vaughn~II, Misra, Popczun, and
  Schaak]{Sines2012}
I.~T. Sines, D.~D. Vaughn~II, R.~Misra, E.~J. Popczun and R.~E. Schaak,
  \emph{J. Solid State Chem.}, 2012, \textbf{196}, 17--20\relax
\mciteBstWouldAddEndPuncttrue
\mciteSetBstMidEndSepPunct{\mcitedefaultmidpunct}
{\mcitedefaultendpunct}{\mcitedefaultseppunct}\relax
\EndOfBibitem
\bibitem[Dutta \emph{et~al.}(2012)Dutta, Maji, Srivastava, Mondal, Biswas,
  Paul, and Adhikary]{Dutta2012}
A.~K. Dutta, S.~K. Maji, D.~N. Srivastava, A.~Mondal, P.~Biswas, P.~Paul and
  B.~Adhikary, \emph{ACS Appl. Mater. \& Interfaces}, 2012, \textbf{4},
  1919--1927\relax
\mciteBstWouldAddEndPuncttrue
\mciteSetBstMidEndSepPunct{\mcitedefaultmidpunct}
{\mcitedefaultendpunct}{\mcitedefaultseppunct}\relax
\EndOfBibitem
\bibitem[Lennie \emph{et~al.}(1995)Lennie, England, and Vaughan]{Lennie1995a}
A.~R. Lennie, K.~E. England and D.~J. Vaughan, \emph{Am. Mineral.}, 1995,
  \textbf{80}, 960--967\relax
\mciteBstWouldAddEndPuncttrue
\mciteSetBstMidEndSepPunct{\mcitedefaultmidpunct}
{\mcitedefaultendpunct}{\mcitedefaultseppunct}\relax
\EndOfBibitem
\bibitem[Csakberenyi-Malasics \emph{et~al.}(2012)Csakberenyi-Malasics,
  Rodriguez-Blanco, Kis, Recnik, Benning, and Posfai]{Malasics2012}
D.~Csakberenyi-Malasics, J.~D. Rodriguez-Blanco, V.~K. Kis, A.~Recnik, L.~G.
  Benning and M.~Posfai, \emph{Chem. Geol.}, 2012, \textbf{294 - 295}, 249 --
  258\relax
\mciteBstWouldAddEndPuncttrue
\mciteSetBstMidEndSepPunct{\mcitedefaultmidpunct}
{\mcitedefaultendpunct}{\mcitedefaultseppunct}\relax
\EndOfBibitem
\bibitem[Bao \emph{et~al.}(2009)Bao, Qui, Huang, Zajdel, Fitzsimmons,
  Zhernenkov, Chang, Fang, Qian, Vehstedt, Yang, Pham, Spinu, and Mao]{Bao2009}
W.~Bao, Y.~Qui, Q.~Huang, M.~A. G. a.~P. Zajdel, M.~R. Fitzsimmons,
  M.~Zhernenkov, S.~Chang, M.~Fang, B.~Qian, E.~K. Vehstedt, J.~Yang, H.~M.
  Pham, L.~Spinu and Z.~Q. Mao, \emph{Phys. Rev. Lett.}, 2009, \textbf{102},
  247001\relax
\mciteBstWouldAddEndPuncttrue
\mciteSetBstMidEndSepPunct{\mcitedefaultmidpunct}
{\mcitedefaultendpunct}{\mcitedefaultseppunct}\relax
\EndOfBibitem
\bibitem[Liu \emph{et~al.}(2010)Liu, Hu, Qian, Fobes, Mao, Bao, Reehuis,
  Kimber, Prokes, Matas, Argyriou, Hiess, Rotaru, Pham, Spinu, Qiu, Thampy,
  Savici, Rodriguez, and Broholm]{Liu2010}
T.~J. Liu, J.~Hu, B.~Qian, D.~Fobes, Z.~Q. Mao, W.~Bao, M.~Reehuis, S.~A.~J.
  Kimber, K.~Prokes, S.~Matas, D.~N. Argyriou, A.~Hiess, A.~Rotaru, H.~Pham,
  L.~Spinu, Y.~Qiu, V.~Thampy, A.~T. Savici, J.~A. Rodriguez and C.~Broholm,
  \emph{Nat. Mater.}, 2010, \textbf{9}, 716--720\relax
\mciteBstWouldAddEndPuncttrue
\mciteSetBstMidEndSepPunct{\mcitedefaultmidpunct}
{\mcitedefaultendpunct}{\mcitedefaultseppunct}\relax
\EndOfBibitem
\bibitem[Bhatia \emph{et~al.}(2011)Bhatia, Rodriguez, Butch, Paglione, and
  Green]{Bhatia2011}
V.~Bhatia, E.~E. Rodriguez, N.~P. Butch, J.~Paglione and M.~A. Green,
  \emph{Chem. Comm.}, 2011, \textbf{47}, 11297\relax
\mciteBstWouldAddEndPuncttrue
\mciteSetBstMidEndSepPunct{\mcitedefaultmidpunct}
{\mcitedefaultendpunct}{\mcitedefaultseppunct}\relax
\EndOfBibitem
\bibitem[Rodriguez \emph{et~al.}(2011)Rodriguez, Stock, Zajdel, Krycka,
  Majkrzak, Zavalij, and Green]{Rodriguez2011_a}
E.~E. Rodriguez, C.~Stock, P.~Zajdel, K.~L. Krycka, C.~F. Majkrzak, P.~Zavalij
  and M.~A. Green, \emph{Phys. Rev. B}, 2011, \textbf{84}, 064403\relax
\mciteBstWouldAddEndPuncttrue
\mciteSetBstMidEndSepPunct{\mcitedefaultmidpunct}
{\mcitedefaultendpunct}{\mcitedefaultseppunct}\relax
\EndOfBibitem
\bibitem[Stock \emph{et~al.}(2011)Stock, Rodriguez, Green, Zavalij, and
  Rodriguez-Rivera]{Stock2011}
C.~Stock, E.~E. Rodriguez, M.~Green, P.~Zavalij and J.~Rodriguez-Rivera,
  \emph{Phys. Rev. B}, 2011, \textbf{84}, 045124\relax
\mciteBstWouldAddEndPuncttrue
\mciteSetBstMidEndSepPunct{\mcitedefaultmidpunct}
{\mcitedefaultendpunct}{\mcitedefaultseppunct}\relax
\EndOfBibitem
\bibitem[Hara \emph{et~al.}(2010)Hara, Takase, Yamasaki, Sato, Miyakawa,
  Umeyama, and Ikeda]{Hara2010}
Y.~Hara, K.~Takase, A.~Yamasaki, H.~Sato, N.~Miyakawa, N.~Umeyama and S.~Ikeda,
  \emph{Physica C}, 2010, \textbf{470}, S313 -- S314\relax
\mciteBstWouldAddEndPuncttrue
\mciteSetBstMidEndSepPunct{\mcitedefaultmidpunct}
{\mcitedefaultendpunct}{\mcitedefaultseppunct}\relax
\EndOfBibitem
\bibitem[B\"ohmer \emph{et~al.}(2015)B\"ohmer, Arai, Hardy, Hattori, Iye, Wolf,
  L\"ohneysen, Ishida, and Meingast]{Boehmer2015}
A.~E. B\"ohmer, T.~Arai, F.~Hardy, T.~Hattori, T.~Iye, T.~Wolf, H.~v.
  L\"ohneysen, K.~Ishida and C.~Meingast, \emph{Phys. Rev. Lett.}, 2015,
  \textbf{114}, 027001\relax
\mciteBstWouldAddEndPuncttrue
\mciteSetBstMidEndSepPunct{\mcitedefaultmidpunct}
{\mcitedefaultendpunct}{\mcitedefaultseppunct}\relax
\EndOfBibitem
\bibitem[Hu \emph{et~al.}(2011)Hu, Cho, Kim, Hodovanets, Straszheim, Tanatar,
  Prozorov, Bud'ko, and Canfield]{Hu2011}
R.~Hu, K.~Cho, H.~Kim, H.~Hodovanets, W.~E. Straszheim, M.~A. Tanatar,
  R.~Prozorov, S.~L. Bud'ko and P.~C. Canfield, \emph{Supercond. Sci. Tech.},
  2011, \textbf{24}, 065006\relax
\mciteBstWouldAddEndPuncttrue
\mciteSetBstMidEndSepPunct{\mcitedefaultmidpunct}
{\mcitedefaultendpunct}{\mcitedefaultseppunct}\relax
\EndOfBibitem
\bibitem[Luo \emph{et~al.}(2011)Luo, Wang, Ying, Yan, Li, Zhang, Wang, Cheng,
  Xiang, Ye, Liu, and Chen]{Luo2011}
X.~G. Luo, X.~F. Wang, J.~J. Ying, Y.~J. Yan, Z.~Y. Li, M.~Zhang, A.~F. Wang,
  P.~Cheng, Z.~J. Xiang, G.~J. Ye, R.~H. Liu and X.~H. Chen, \emph{New J.
  Phys.}, 2011, \textbf{13}, 053011\relax
\mciteBstWouldAddEndPuncttrue
\mciteSetBstMidEndSepPunct{\mcitedefaultmidpunct}
{\mcitedefaultendpunct}{\mcitedefaultseppunct}\relax
\EndOfBibitem
\bibitem[Lei \emph{et~al.}(2011)Lei, Abeykoon, Bozin, and Petrovic]{Lei2011}
H.~Lei, M.~Abeykoon, E.~S. Bozin and C.~Petrovic, \emph{Phys. Rev. B}, 2011,
  \textbf{83}, 180503\relax
\mciteBstWouldAddEndPuncttrue
\mciteSetBstMidEndSepPunct{\mcitedefaultmidpunct}
{\mcitedefaultendpunct}{\mcitedefaultseppunct}\relax
\EndOfBibitem
\bibitem[McQueen \emph{et~al.}(2009)McQueen, Huang, Ksenofontov, Felser, Xu,
  Zandbergen, Hor, Allred, Williams, Qu, Checkelsky, Ong, and
  Cava]{McQueen2009}
T.~M. McQueen, Q.~Huang, V.~Ksenofontov, C.~Felser, Q.~Xu, H.~Zandbergen, Y.~S.
  Hor, J.~Allred, A.~J. Williams, D.~Qu, J.~Checkelsky, N.~P. Ong and R.~J.
  Cava, \emph{Phys. Rev. B}, 2009, \textbf{79}, 014522\relax
\mciteBstWouldAddEndPuncttrue
\mciteSetBstMidEndSepPunct{\mcitedefaultmidpunct}
{\mcitedefaultendpunct}{\mcitedefaultseppunct}\relax
\EndOfBibitem
\bibitem[Sun \emph{et~al.}(2015)Sun, Woodruff, Cassidy, Allcroft, Sedlmaier,
  Thompson, Bingham, Forder, Cartenet, Mary, Ramos, Foronda, Williams, Li,
  Blundell, and Clarke]{Sun2015}
H.~Sun, D.~N. Woodruff, S.~J. Cassidy, G.~M. Allcroft, S.~J. Sedlmaier, A.~L.
  Thompson, P.~A. Bingham, S.~D. Forder, S.~Cartenet, N.~Mary, S.~Ramos, F.~R.
  Foronda, B.~H. Williams, X.~Li, S.~J. Blundell and S.~J. Clarke,
  \emph{Inorganic Chemistry}, 2015, \textbf{54}, 1958--1964\relax
\mciteBstWouldAddEndPuncttrue
\mciteSetBstMidEndSepPunct{\mcitedefaultmidpunct}
{\mcitedefaultendpunct}{\mcitedefaultseppunct}\relax
\EndOfBibitem
\bibitem[Smith and Chu(1967)]{Smith1967}
T.~F. Smith and C.~W. Chu, \emph{Phys. Rev.}, 1967, \textbf{159},
  353--358\relax
\mciteBstWouldAddEndPuncttrue
\mciteSetBstMidEndSepPunct{\mcitedefaultmidpunct}
{\mcitedefaultendpunct}{\mcitedefaultseppunct}\relax
\EndOfBibitem
\bibitem[Lei \emph{et~al.}(2011)Lei, Abeykoon, Bozin, Wang, Warren, and
  Petrovic]{Lei2011a}
H.~Lei, M.~Abeykoon, E.~S. Bozin, K.~Wang, J.~B. Warren and C.~Petrovic,
  \emph{Phys. Rev. Lett.}, 2011, \textbf{107}, 137002\relax
\mciteBstWouldAddEndPuncttrue
\mciteSetBstMidEndSepPunct{\mcitedefaultmidpunct}
{\mcitedefaultendpunct}{\mcitedefaultseppunct}\relax
\EndOfBibitem
\bibitem[Shoemaker \emph{et~al.}(2012)Shoemaker, Chung, Claus, Francisco, Avci,
  Llobet, and Kanatzidis]{Shoemaker2012}
D.~P. Shoemaker, D.~Y. Chung, H.~Claus, M.~C. Francisco, S.~Avci, A.~Llobet and
  M.~G. Kanatzidis, \emph{Phys. Rev. B}, 2012, \textbf{86}, 184511\relax
\mciteBstWouldAddEndPuncttrue
\mciteSetBstMidEndSepPunct{\mcitedefaultmidpunct}
{\mcitedefaultendpunct}{\mcitedefaultseppunct}\relax
\EndOfBibitem
\bibitem[Neilson and McQueen(2012)]{Neilson2012}
J.~R. Neilson and T.~M. McQueen, \emph{J. Amer. Chem. Soc.}, 2012,
  \textbf{134}, 7750--7757\relax
\mciteBstWouldAddEndPuncttrue
\mciteSetBstMidEndSepPunct{\mcitedefaultmidpunct}
{\mcitedefaultendpunct}{\mcitedefaultseppunct}\relax
\EndOfBibitem
\bibitem[Zhou \emph{et~al.}()Zhou, Borg, Lynn, Saha, Paglione, and
  Rodriguez]{Zhou2015}
X.~Zhou, C.~K.~H. Borg, J.~W. Lynn, S.~R. Saha, J.~Paglione and E.~E.
  Rodriguez, \emph{http://arxiv.org/abs/1512.03399}\relax
\mciteBstWouldAddEndPuncttrue
\mciteSetBstMidEndSepPunct{\mcitedefaultmidpunct}
{\mcitedefaultendpunct}{\mcitedefaultseppunct}\relax
\EndOfBibitem
\bibitem[Dong \emph{et~al.}(2015)Dong, Jin, Yuan, Zhou, Yuan, Huang, Hua, Sun,
  Zheng, Hu, Mao, Ma, Zhang, Zhou, and Zhao]{Dong2015}
X.~Dong, K.~Jin, D.~Yuan, H.~Zhou, J.~Yuan, Y.~Huang, W.~Hua, J.~Sun, P.~Zheng,
  W.~Hu, Y.~Mao, M.~Ma, G.~Zhang, F.~Zhou and Z.~Zhao, \emph{Phys. Rev. B},
  2015, \textbf{92}, 064515\relax
\mciteBstWouldAddEndPuncttrue
\mciteSetBstMidEndSepPunct{\mcitedefaultmidpunct}
{\mcitedefaultendpunct}{\mcitedefaultseppunct}\relax
\EndOfBibitem
\bibitem[Pachmayr \emph{et~al.}(2016)Pachmayr, Fehn, and
  Johrendt]{Pachmayr2016}
U.~Pachmayr, N.~Fehn and D.~Johrendt, \emph{Chem. Commun.}, 2016,  --\relax
\mciteBstWouldAddEndPuncttrue
\mciteSetBstMidEndSepPunct{\mcitedefaultmidpunct}
{\mcitedefaultendpunct}{\mcitedefaultseppunct}\relax
\EndOfBibitem
\bibitem[Margadonna \emph{et~al.}(2009)Margadonna, Takabayashi, Ohishi,
  Mizuguchi, Takano, Kagayama, Nakagawa, Takata, and Prassides]{Margadonna2009}
S.~Margadonna, Y.~Takabayashi, Y.~Ohishi, Y.~Mizuguchi, Y.~Takano, T.~Kagayama,
  T.~Nakagawa, M.~Takata and K.~Prassides, \emph{Phys. Rev. B}, 2009,
  \textbf{80}, 064506\relax
\mciteBstWouldAddEndPuncttrue
\mciteSetBstMidEndSepPunct{\mcitedefaultmidpunct}
{\mcitedefaultendpunct}{\mcitedefaultseppunct}\relax
\EndOfBibitem
\bibitem[Rodriguez \emph{et~al.}(2013)Rodriguez, Sokolov, Stock, Green,
  Sobolev, Rodriguez-Rivera, Cao, and Daoud-Aladine]{Rodriguez2013}
E.~E. Rodriguez, D.~A. Sokolov, C.~Stock, M.~A. Green, O.~Sobolev, J.~A.
  Rodriguez-Rivera, H.~Cao and A.~Daoud-Aladine, \emph{Physical Review B},
  2013, \textbf{88}, 165110\relax
\mciteBstWouldAddEndPuncttrue
\mciteSetBstMidEndSepPunct{\mcitedefaultmidpunct}
{\mcitedefaultendpunct}{\mcitedefaultseppunct}\relax
\EndOfBibitem
\bibitem[Lennie \emph{et~al.}(1997)Lennie, Redfern, Champness, Stoddart,
  Schofield, and Vaughan]{Lennie1997}
A.~R. Lennie, S.~A.~T. Redfern, P.~E. Champness, C.~P. Stoddart, P.~F.
  Schofield and D.~J. Vaughan, \emph{Am. Mineral.}, 1997, \textbf{82},
  302--309\relax
\mciteBstWouldAddEndPuncttrue
\mciteSetBstMidEndSepPunct{\mcitedefaultmidpunct}
{\mcitedefaultendpunct}{\mcitedefaultseppunct}\relax
\EndOfBibitem
\bibitem[Werthamer \emph{et~al.}(1966)Werthamer, Helfand, and
  Hohenberg]{Werthamer1966}
N.~R. Werthamer, E.~Helfand and P.~C. Hohenberg, \emph{Phys. Rev.}, 1966,
  \textbf{147}, 295--302\relax
\mciteBstWouldAddEndPuncttrue
\mciteSetBstMidEndSepPunct{\mcitedefaultmidpunct}
{\mcitedefaultendpunct}{\mcitedefaultseppunct}\relax
\EndOfBibitem
\bibitem[Tinkham(1963)]{Tinkham1963}
M.~Tinkham, \emph{Phys. Rev.}, 1963, \textbf{129}, 2413--2422\relax
\mciteBstWouldAddEndPuncttrue
\mciteSetBstMidEndSepPunct{\mcitedefaultmidpunct}
{\mcitedefaultendpunct}{\mcitedefaultseppunct}\relax
\EndOfBibitem
\bibitem[Ketterson and Song(1999)]{Ketterson1999}
J.~Ketterson and S.~Song, \emph{Superconductivity}, Cambridge University Press,
  40 West 20th Street, New York, NY 10011-4211, USA, 1999, pp. 45--46\relax
\mciteBstWouldAddEndPuncttrue
\mciteSetBstMidEndSepPunct{\mcitedefaultmidpunct}
{\mcitedefaultendpunct}{\mcitedefaultseppunct}\relax
\EndOfBibitem
\bibitem[Hsu \emph{et~al.}(2008)Hsu, Luo, Yeh, Chen, Huang, Wu, Lee, Huang,
  Chu, Yan,\emph{et~al.}]{Hsu2008}
F.-C. Hsu, J.-Y. Luo, K.-W. Yeh, T.-K. Chen, T.-W. Huang, P.~M. Wu, Y.-C. Lee,
  Y.-L. Huang, Y.-Y. Chu, D.-C. Yan \emph{et~al.}, \emph{P. Natl. Acad. Sci.
  USA}, 2008, \textbf{105}, 14262--14264\relax
\mciteBstWouldAddEndPuncttrue
\mciteSetBstMidEndSepPunct{\mcitedefaultmidpunct}
{\mcitedefaultendpunct}{\mcitedefaultseppunct}\relax
\EndOfBibitem
\bibitem[Hu \emph{et~al.}(2009)Hu, Bozin, Warren, and Petrovic]{Hu2009}
R.~Hu, E.~S. Bozin, J.~B. Warren and C.~Petrovic, \emph{Phys. Rev. B}, 2009,
  \textbf{80}, 214514\relax
\mciteBstWouldAddEndPuncttrue
\mciteSetBstMidEndSepPunct{\mcitedefaultmidpunct}
{\mcitedefaultendpunct}{\mcitedefaultseppunct}\relax
\EndOfBibitem
\bibitem[Abdel-Hafiez \emph{et~al.}(2015)Abdel-Hafiez, Zhang, Cao, Duan,
  Karapetrov, Pudalov, Vlasenko, Sadakov, Knyazev, Romanova, Chareev, Volkova,
  Vasiliev, and Chen]{Abdel-Hafiez2015}
M.~Abdel-Hafiez, Y.-Y. Zhang, Z.-Y. Cao, C.-G. Duan, G.~Karapetrov, V.~M.
  Pudalov, V.~A. Vlasenko, A.~V. Sadakov, D.~A. Knyazev, T.~A. Romanova, D.~A.
  Chareev, O.~S. Volkova, A.~N. Vasiliev and X.-J. Chen, \emph{Phys. Rev. B},
  2015, \textbf{91}, 165109\relax
\mciteBstWouldAddEndPuncttrue
\mciteSetBstMidEndSepPunct{\mcitedefaultmidpunct}
{\mcitedefaultendpunct}{\mcitedefaultseppunct}\relax
\EndOfBibitem
\bibitem[Millis \emph{et~al.}(1988)Millis, Sachdev, and Varma]{Millis1988}
A.~J. Millis, S.~Sachdev and C.~M. Varma, \emph{Phys. Rev. B.}, 1988,
  \textbf{37}, 4795\relax
\mciteBstWouldAddEndPuncttrue
\mciteSetBstMidEndSepPunct{\mcitedefaultmidpunct}
{\mcitedefaultendpunct}{\mcitedefaultseppunct}\relax
\EndOfBibitem
\bibitem[Vaughan and Ridout(1971)]{Vaughan1971}
D.~Vaughan and M.~Ridout, \emph{J. Inorg. Nucl. Chem.}, 1971, \textbf{33}, 741
  -- 746\relax
\mciteBstWouldAddEndPuncttrue
\mciteSetBstMidEndSepPunct{\mcitedefaultmidpunct}
{\mcitedefaultendpunct}{\mcitedefaultseppunct}\relax
\EndOfBibitem
\bibitem[Subedi \emph{et~al.}(2008)Subedi, Zhang, Singh, and Du]{Subedi2008}
A.~Subedi, L.~Zhang, D.~J. Singh and M.~H. Du, \emph{Phys. Rev. B}, 2008,
  \textbf{78}, 134514\relax
\mciteBstWouldAddEndPuncttrue
\mciteSetBstMidEndSepPunct{\mcitedefaultmidpunct}
{\mcitedefaultendpunct}{\mcitedefaultseppunct}\relax
\EndOfBibitem
\bibitem[Kwon \emph{et~al.}(2011)Kwon, Refson, Bone, Qiao, Yang, Liu, and
  Sposito]{Kwon2011}
K.~D. Kwon, K.~Refson, S.~Bone, R.~Qiao, W.-l. Yang, Z.~Liu and G.~Sposito,
  \emph{Phys. Rev. B}, 2011, \textbf{83}, 064402\relax
\mciteBstWouldAddEndPuncttrue
\mciteSetBstMidEndSepPunct{\mcitedefaultmidpunct}
{\mcitedefaultendpunct}{\mcitedefaultseppunct}\relax
\EndOfBibitem
\bibitem[Devey \emph{et~al.}(2008)Devey, Grau-Crespo, and de~Leeuw]{Devey2008}
A.~J. Devey, R.~Grau-Crespo and N.~H. de~Leeuw, \emph{J. Phys. Chem. C}, 2008,
  \textbf{112}, 10960--10967\relax
\mciteBstWouldAddEndPuncttrue
\mciteSetBstMidEndSepPunct{\mcitedefaultmidpunct}
{\mcitedefaultendpunct}{\mcitedefaultseppunct}\relax
\EndOfBibitem
\bibitem[Brgoch and Miller(2012)]{Brgoch2012}
J.~Brgoch and G.~J. Miller, \emph{J. Phys. Chem. A}, 2012, \textbf{116},
  2234--2243\relax
\mciteBstWouldAddEndPuncttrue
\mciteSetBstMidEndSepPunct{\mcitedefaultmidpunct}
{\mcitedefaultendpunct}{\mcitedefaultseppunct}\relax
\EndOfBibitem
\bibitem[Greenfield \emph{et~al.}(2015)Greenfield, Kamali, Lee, and
  Kovnir]{Greenfield2015}
J.~T. Greenfield, S.~Kamali, K.~Lee and K.~Kovnir, \emph{Chem. Mater.}, 2015,
  \textbf{27}, 588--596\relax
\mciteBstWouldAddEndPuncttrue
\mciteSetBstMidEndSepPunct{\mcitedefaultmidpunct}
{\mcitedefaultendpunct}{\mcitedefaultseppunct}\relax
\EndOfBibitem
\bibitem[Lee \emph{et~al.}(2012)Lee, Kihou, Iyo, Kito, Shirage, and
  Eisaki]{Lee2012}
C.~Lee, K.~Kihou, A.~Iyo, H.~Kito, P.~Shirage and H.~Eisaki, \emph{Solid State
  Commun.}, 2012, \textbf{152}, 644 -- 648\relax
\mciteBstWouldAddEndPuncttrue
\mciteSetBstMidEndSepPunct{\mcitedefaultmidpunct}
{\mcitedefaultendpunct}{\mcitedefaultseppunct}\relax
\EndOfBibitem
\bibitem[Mizuguchi \emph{et~al.}(2010)Mizuguchi, Hara, Deguchi, Tsuda,
  Yamaguchi, Takeda, Kotegawa, Tou, and Takano]{Mizuguchi2010}
Y.~Mizuguchi, Y.~Hara, K.~Deguchi, S.~Tsuda, T.~Yamaguchi, K.~Takeda,
  H.~Kotegawa, H.~Tou and Y.~Takano, \emph{Supercond. Sci. Tech.}, 2010,
  \textbf{23}, 054013\relax
\mciteBstWouldAddEndPuncttrue
\mciteSetBstMidEndSepPunct{\mcitedefaultmidpunct}
{\mcitedefaultendpunct}{\mcitedefaultseppunct}\relax
\EndOfBibitem
\bibitem[Imai \emph{et~al.}(2009)Imai, Ahilan, Ning, McQueen, and
  Cava]{Imai2009}
T.~Imai, K.~Ahilan, F.~L. Ning, T.~M. McQueen and R.~J. Cava, \emph{Phys. Rev.
  Lett.}, 2009, \textbf{102}, 177005\relax
\mciteBstWouldAddEndPuncttrue
\mciteSetBstMidEndSepPunct{\mcitedefaultmidpunct}
{\mcitedefaultendpunct}{\mcitedefaultseppunct}\relax
\EndOfBibitem
\bibitem[Mizuguchi \emph{et~al.}(2009)Mizuguchi, Tomioka, Tsuda, Yamaguchi, and
  Takano]{Mizuguchi2009}
Y.~Mizuguchi, F.~Tomioka, S.~Tsuda, T.~Yamaguchi and Y.~Takano, \emph{Appl.
  Phys. Lett.}, 2009, \textbf{94}, --\relax
\mciteBstWouldAddEndPuncttrue
\mciteSetBstMidEndSepPunct{\mcitedefaultmidpunct}
{\mcitedefaultendpunct}{\mcitedefaultseppunct}\relax
\EndOfBibitem
\bibitem[Zajdel \emph{et~al.}(2010)Zajdel, Hsieh, Rodriguez, Butch, Magill,
  Paglione, Zavalij, Suchomel, and Green]{Zajdel2010}
P.~Zajdel, P.-Y. Hsieh, E.~E. Rodriguez, N.~P. Butch, J.~D. Magill,
  J.~Paglione, P.~Zavalij, M.~R. Suchomel and M.~A. Green, \emph{J. Amer. Chem.
  Soc.}, 2010, \textbf{132}, 13000--13007\relax
\mciteBstWouldAddEndPuncttrue
\mciteSetBstMidEndSepPunct{\mcitedefaultmidpunct}
{\mcitedefaultendpunct}{\mcitedefaultseppunct}\relax
\EndOfBibitem
\end{mcitethebibliography}
\providecommand*{\mcitethebibliography}{\thebibliography}
\csname @ifundefined\endcsname{endmcitethebibliography}
{\let\endmcitethebibliography\endthebibliography}{}

\end{document}